%% file: main.tex
\documentclass[10pt,journal,compsoc]{IEEEtran}
\ifCLASSOPTIONcompsoc
  \usepackage[nocompress]{cite}
\else
  \usepackage{cite}
\fi
\usepackage{hyperref}
\usepackage{gensymb}
\usepackage{cancel}
\usepackage{mathtools}
\usepackage{latexsym}
\usepackage{booktabs}
\usepackage{amsmath,bm}
\usepackage{wrapfig}
\usepackage{lineno}
\usepackage{amsthm}
\usepackage{psfrag}
\usepackage{mathtools}
\usepackage{floatflt}
\usepackage{subfigure}
\usepackage{cases, enumerate}
\usepackage{threeparttable}
\usepackage{algorithm}
\usepackage{algpseudocode}
\usepackage{tabularx}
\usepackage{array}
\usepackage{amssymb}
\usepackage{pifont}

\usepackage{graphicx}
\newcommand{\cmark}{\ding{51}}%
\newcommand{\xmark}{\ding{55}}%
\usepackage{xcolor}

\title{Penetration-free Solid-Fluid Interaction on Shells and Rods}
\author{Yuchen Sun*, Jinyuan Liu*, Yin Yang, Chenfanfu Jiang, Minchen Li, Bo Zhu\thanks{The main body of this work was completed as part of Jinyuan Liu's Ph.D. dissertation at Dartmouth College in June 2024 \cite{liu2024crossscale}}}

\input{0-abstract}
\begin{document}
\maketitle
\input{1-introduction}

\input{2-related_work}
\input{3-physical_model}
\input{4-numerical_algorithm}
\input{5-results}

\bibliographystyle{IEEEtran}
\bibliography{ref}

\input{author_info}
\end{document}

%% file: 0-abstract.tex
\IEEEtitleabstractindextext{%
\begin{abstract}
We introduce a novel approach to simulate the interaction between fluids and thin elastic solids without any penetration. Our approach is centered around an optimization system augmented with barriers, which aims to find a configuration that ensures the absence of penetration while enforcing incompressibility for the fluids and minimizing elastic potentials for the solids. Unlike previous methods that primarily focus on velocity coherence at the fluid-solid interfaces, we demonstrate the effectiveness and flexibility of explicitly resolving positional constraints, including both explicit representation of solid positions and the implicit representation of fluid level-set interface. To preserve the volume of the fluid, we propose a simple yet efficient approach that adjusts the associated level-set values. Additionally, we develop a distance metric capable of measuring the separation between an implicitly represented surface and a Lagrangian object of arbitrary codimension. By integrating the inertia, solid elastic potential, damping, barrier potential, and fluid incompressibility within a unified system, we are able to robustly simulate a wide range of processes involving fluid interactions with lower-dimensional objects such as shells and rods. These processes include topology changes, bouncing, splashing, sliding, rolling, floating, and more.
\end{abstract}

}

%% file: 1-introduction.tex
\section{Introduction}

The interactions between interfacial fluids and thin solids, such as shells and rods, are widespread phenomena that occur frequently in nature. These interactions manifest in various captivating forms, including raindrops falling on lotus leaves, dewdrops gathering on grasses, sweat drops coalescing between wet hairs, aquatic insects skimming across water surfaces, and rain splashes pouring onto an umbrella surface, just to name a few examples.
These visually intriguing processes often involve vivid vibrations of thin shells or rods, as well as deformation and splashing of liquid volumes due to surface tension. More interestingly, these phenomena also encompass a myriad of contact interactions between the solid surface and fluid interface, which give rise to various visually appealing interaction behaviors on small scales.
These solid-fluid coupling systems have consistently captured the attention of researchers across computer graphics and computational physics, motivating the development of a plethora of numerical algorithms (e.g.,\cite{ruan2021solid, takahashi2022elastomonolith, robinson2008two}) to unravel their geometric and dynamic complexities in different aspects.

The robust treatment of solid-fluid contact is fundamental in constructing an effective numerical solver to model these intricate phenomena. The interaction between interfacial fluid volumes and thin elastic objects primarily relies on contact to exchange momentum and enable dynamic coupling. In scenarios involving surface tension, the elastocapillary coupling mechanics also rely on the evolution of the contact area and its contours, which drive the interfacial energy minimization processes. Therefore, the development of an effective procedure capable of capturing and representing the solid-fluid contact process plays a critical role in facilitating these challenging simulations.

However, achieving robust contact handling between interfacial fluids and thin objects presents significant challenges. These challenges can be categorized into two key aspects: \emph{capturing contacts} and \emph{rigorously enforcing constraints}.
Firstly, capturing the contact process between the fluid interface and solid surfaces is difficult. Existing approaches often struggle when dealing with thin objects like shells and rods. The small volumes of these thin Lagrangian objects make them somewhat "invisible" to the Eulerian grid. Consequently, additional algorithmic steps, such as ray casting \cite{guendelman2005coupling}, are necessary to detect and explicitly incorporate these thin volumes into the coupling system. However, these processes can be discontinuous, posing difficulties in accommodating coupling system (e.g., \cite{batty2007fast}). 
Secondly, enforcing constraints during solid-fluid contact is challenging. Unlike solid-solid interaction problems, where positions can be used directly as variables to enforce non-penetration contact, traditional solid-fluid coupling systems solve for fluid velocities. Although the velocities of fluids and solids may perfectly match at their contacts, the advected fluid interface and the updated solid surface can still exhibit penetration due to the inherent nonlinearity of level-set advection. Techniques such as cut-cell \cite{zarifi2017positive}, enhanced level-set method \cite{enright2002hybrid}, and high-order advection schemes \cite{selle2008unconditionally} can mitigate these penetration issues to some extent, but they do not fully resolve the problem. Similar issues also occur for rigorously enforcing incompressibility of the fluid volume. If not properly addressed, these penetration and compressibility issues may lead to instability and inaccuracy in solid-fluid coupling simulations, resulting in visual artifacts such as leakage, ghost forces, and volume changes. These artifacts are particularly noticeable in coupling systems with thin structures.

To overcome these challenges, we propose a position-level framework that addresses contacts between fluid interfaces and thin elastic objects. Our approach focuses on rigorously enforcing non-penetration and incompressible volume constraints during solid-fluid contact through an optimization process.
At the core of our model is a novel implicit distance metric derived from the level-set field. This metric measures the relative distance between a solid vertex and its closest fluid surface patch. It is then formulated as a barrier potential to enforce interpenetration-free contacts.
Our framework seamlessly integrates inertia, fluid incompressibility, solid elasticity, physical damping, collision, and contact into a unified system. 
By incorporating these advancements, our framework provides a comprehensive and versatile solution for simulating solid-fluid coupling phenomena with thin objects, opening up new possibilities for understanding and visualizing complex interfacial dynamics.

Our solid-fluid interaction system draws inspiration from two categories of established work: solid-fluid coupling schemes \cite{robinson2008two} and incremental potential contact (IPC) \cite{li2020incremental}, which have demonstrated significant success in handling solid-fluid and solid-solid coupling problems with diverse geometries and physics.
In our framework, we combine the computational advantages of both approaches to establish a coupling systems. This system is built upon implicit interface positions enhanced by novel constraint enforcement, Newton solver, and time integration schemes, allowing us to efficiently address the coupling optimization problem and facilitate solid-fluid contact treatments. In contrast to fragmented approaches that treat different interaction targets separately, our framework is based on a unified energy-based formulation for solid-fluid interaction.

We summarize the main contributions of our approach as:
\begin{itemize}
    \item A position-level contact model between solids and fluids equipped with conservation-of-volume and penetration-free constraints;
    \item A novel approach to enforce the incompressibility of the level-set method, without the need for high-order advection schemes or auxiliary particles;
    \item A Newton solver with line search filtered by Continuous Collision Detection (CCD) to resolve positional constraints between solids and fluids;
    \item A unified framework to simulate interactions between Eulerian fluids and Lagrangian solids of arbitrary codimension.
\end{itemize}

%% file: 2-related_work.tex
\section{Related Work}
\subsection{Interfacial Flow}
Simulating incompressible liquids has received long-lasting attention in computer graphics for its visually intriguing interfacial dynamics, continuous topological changes, and contrasting multi-scale flow details (see \cite{zsolnai2022flow} for an overview). Grid-based approaches typically employ the level-set method \cite{sethian2003level} for surface tracking, and great efforts have been put into improving the simulation quality by introducing auxiliary particles \cite{enright2002hybrid}, mitigating dissipation through high-order schemes \cite{selle2008unconditionally}, enhancing vorticity conservation \cite{fedkiw2001visual,yang2021clebsch}, and improving scalability \cite{mcadams2010parallel} and adaptivity \cite{setaluri2014spgrid}. Other popular methods are based on meshes \cite{wojtan2011liquid}, particles \cite{macklin2013position, ihmsen2014sph}, and hybrid Eulerian-Lagrangian models \cite{jiang2015affine,zhu2005animating}. Similarly, stable and efficient modeling of surface tension effects has been explored on grids \cite{zheng2009simulation,liu2022hydrophobic}, simplicial complexes \cite{zhu2014codimensional,da2016surface}, particles \cite{wang2021thin,xing2022position}, and hybrids \cite{zheng2015new,hyde2020implicit}. Combining free surface flow with surface tension has triggered the animations of a range of physical phenomena, including blowing and bursting of bubbles \cite{wang2020codimensional,wang2021thin}, floating of heavy objects \cite{ruan2021solid,liu2022hydrophobic}, coalescence \cite{fei2017multi}, permeation \cite{fei2018multi}, milk crown \cite{zheng2015new}, and splashing \cite{xing2022position}. At the center of all these researches lies the incompressibility of fluid, which typically follows the projection concept. However, enforcing such constraint on particle-based fluids usually has stability and convergence issues \cite{monaghan2005smooth},. For grid-based fluids, the level-set method also cannot preserve volume by nature, especially at regions of high curvature, narrow width and close to the boundary, where noticeable volume gain or loss can happen. 

\subsection{Solid-Fluid Interaction}
Coupling between solids and fluids can be modeled from different perspectives, such as pure Eulerian \cite{teng2016eulerian}, pure Lagrangian \cite{akinci2012versatile,xie2023contact}, and their hybrids \cite{guendelman2005coupling,hu2018moving,fang2020iq}. 

A number of works couple Eulerian fluids with point-like or granular entities, mostly through drag-type momentum exchange. One approach is the multi-species mixture model where porous sand grains exchange momentum with water through drag and Darcy-like interactions \cite{tampubolon2017multi}. A hybrid Lagrangian–Eulerian scheme is proposed to couple bubble particles to liquid flows using drag and pressure forces \cite{patkar2013hybrid}.  Recently Loki \cite{lesser2022loki}, developed as a production-oriented multiphysics system, treats granular particles, rigid bodies, and fluids in a unified solver with drag and pressure driven coupling mechanisms.

Monolithic two-way coupling approaches between Eulerian fluids and Lagrangian solids have shown superb accuracy, robustness and flexibility, suitable for a vast range of solid types including rigids \cite{batty2007fast}, deformables \cite{robinson2008two, hyde2019unified}, fabrics \cite{fei2018multi}, rods \cite{fei2017multi}, reduced elastic solids \cite{lu2016two}, and articulated bodies \cite{ruan2021solid}. In order to properly accommodate different kinds of interactive forces, a variational formulation on pressure was introduced by \cite{batty2007fast}, and quickly extended to handle elasticity \cite{zarifi2017positive}, damping \cite{robinson2011symmetric}, viscosity \cite{takahashi2019geometrically}, friction \cite{narain2010free}, surface tension \cite{ruan2021solid}, elastocapillarity \cite{liu2022hydrophobic}, and frictional contacts \cite{takahashi2022elastomonolith}.
While these methods only enforce equivalence of the normal velocity component to prevent interpenetration, our approach enforces non-penetration directly at the positional level. In addition, interactions between fluids and volumetric solids cannot be directly applied to lower dimensional objects most of the time. Special treatments are usually required to avoid unnatural behaviors such as leakage when dealing with thin shells or rods. For instance, \cite{guendelman2005coupling} developed a robust one-sided interpolation via ray casting against thickened triangle wedges. \cite{fei2018multi} adapted the affine particle-in-cell method for fluids, and introduced additional physical models based on mixture theory. \cite{ruan2021solid} manually projected contact particles to the desired side of the thin shell in an unphysical manner. A fundamental coupling system that can seamlessly handle elastic solids of arbitrary codimension is still missing and remains a challenging task.

\subsection{Contact Handling}
Building physically plausible contact models has been extensively investigated in mechanics, robotics, and computer graphics \cite{andrews2022contact,brogliato1999nonsmooth}. Following \cite{baraff1991coping}, rigid-body contacts are often formulated as constraint-based linear complementarity problems (LCP) \cite{stewart2000implicit,kaufman2008staggered,baraff1994fast} solved with projected Gauss-Seidel (PGS). A similar formulation is by using the quadratic programming (QP) model \cite{redon2002fast,renouf2005conjugate,macklin2020primal}, which can benefit from more flexible and efficient solving techniques, such as projected gradient descent \cite{mazhar2015using} and the augmented Lagrangian method \cite{takahashi2021frictionalmonolith}. For elastic solids, LCP-based and QP-based contact handling formulations are also popular, supplemented with penalty-based methods on resolving contact forces \cite{bridson2002robust,tang2012continuous,xu2014implicit}. While traditional methods typically enforce contact via approximate constraints based on signed distances or volume overlap, recent approaches derived from the incremental potential contact (IPC) model \cite{li2020incremental} have demonstrated substantially improved robustness and accuracy by relying on geometrically exact unsigned distances. IPC formulates contact as a conservative barrier potential embedded within an incremental energy minimization framework, converting non-penetration and frictional interactions into smooth, differentiable terms that integrate seamlessly with standard implicit time stepping. Built on this formulation, IPC has been extended to rigid bodies \cite{ferguson2021intersection}, codimensional solids \cite{li2020codimensional}, reduced deformable objects \cite{lan2021medial}, stiff materials \cite{lan2022affine}, embedded interfaces \cite{zhao2022barrier}, hybrid multibody systems \cite{chen2022unified}, FEM-SPH coupled domains \cite{xie2023contact} FEM-MPM coupled systems \cite{xuan2024duo}. However, this idea has not yet been extended to Eulerian fluid formulations. In contrast, our method accommodates the coupling between Lagrangian solids and Eulerian fluids within a unified framework.

%% file: 3-physical_model.tex
\section{Physical Model}
\begin{wrapfigure}{r}{4.1cm}
\centering
\includegraphics[width=4cm]{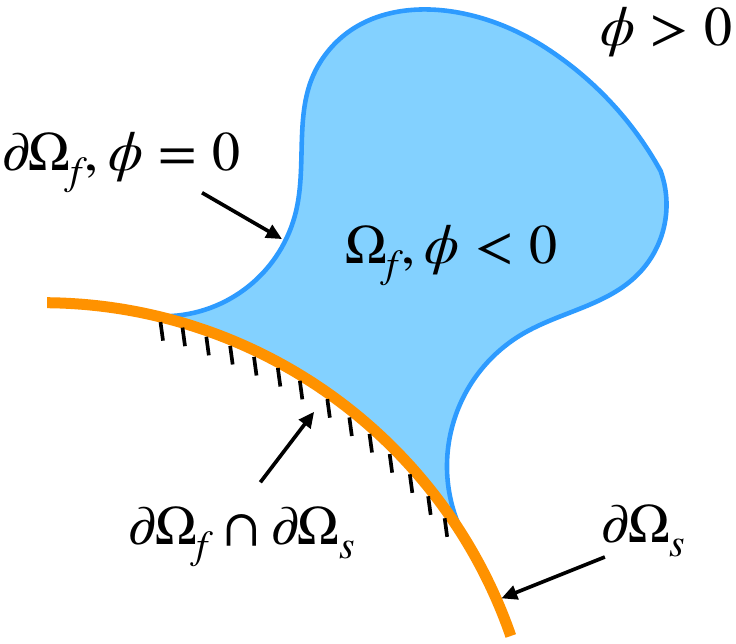}
\caption{Illustration of the geometric setup}
\label{fig:mark_cell}
\end{wrapfigure}
\textbf{Geometric domain.}
We first define the geometric domain of our solid-fluid coupling problem. As shown in the inset figure, we denote the fluid domain as $\Omega_f$ and the solid domain as $\Omega_s$, with their boundaries specified as $\partial \Omega_f$ and $\partial \Omega_s$, respectively. The fluid domain is implicitly specified as a signed distance field $\phi$ such that $\phi<0$ indicates the fluid volume's interior and $\phi>0$ indicates the exterior.The fluid interface is the boundary of the fluid domain where $\phi=0$. The contact between solid and fluid is simply $\partial \Omega_f \cap \partial \Omega_s$ i.e., all the points on the solid boundary where $\phi=0$. 

\textbf{Elastic solid.}
The dynamics of an elastic solid can be represented by the Lagrangian form of the momentum equation:
\begin{equation}
    \rho_s\bm a=\nabla^{\bm X} \cdot \bm{P}+\rho_s\bm g,
    \label{eq:elastodynamics}
\end{equation}
where $\rho_s$ is the initial solid density, $\bm a$ is the acceleration, and $\bm X$ is the material (or initial) space coordinate. $\bm P$ is the first Piola-Kirchhoff stress tensor, and $\bm g$ is the gravity acceleration. In this work, we focus on two types of elastic objects (shells and rods), whose energy forms and derivatives have been thoroughly investigated in the previous literature (e.g., see \cite{grinspun2003discrete,bergou2008discrete}). 

\textbf{Interfacial fluid.}
We define the motion of an inviscid fluid using the incompressible Euler equations:
\begin{equation}
\begin{dcases}
    \frac{\partial \bm{u}}{\partial t} + (\bm{u}\cdot \nabla) \bm{u} = -\frac{1}{\rho}\nabla p + \bm{g},\\
    \nabla \cdot \bm{u} = 0,
    \label{eq:fluid_ns}
\end{dcases}
\end{equation}
where $\bm u$ is the velocity vector field; $\bm g$ is the gravitational acceleration; and $p$ and $\rho$ are fluid pressure and density. 

The momentum equation of Equation~\ref{eq:fluid_ns} can also be expressed in terms of the Cauchy stress tensor $\bm \sigma$ as:
\begin{equation}
    \rho\frac{\mathrm{D} \bm u}{\mathrm{D} t} =\nabla\cdot\bm\sigma + \rho\bm g
\end{equation}
where $\bm \sigma = -p\bm I$, $\frac{\mathrm{D} \bm u}{\mathrm{D} t}$ is the full derivative of velocity with respect to time.

Equation~\ref{eq:fluid_ns} are subject to two sets of boundary conditions: solid boundary conditions and free-surface boundary conditions. 
The solid boundary conditions capture the contact constraints between the fluid and solid domains, while the free-surface boundary conditions handle the effects of surface tension.
We discuss each of these boundary conditions below. 

At the contact interface between the fluid and solid domains i.e., $\partial \Omega_f \cap \partial \Omega_s$, it is required that 1) there should be no interpenetration between the fluid and solid volumes at the positional level, and 2) the Neumann boundary conditions are activated at the velocity level such that fluid and solid move under the same velocity along the contact normal. Mathematically, those two constraints can be summarized as:

\begin{equation}
\begin{dcases}
    d(\partial \Omega_f, \bm x)=\phi(\bm x)\geq 0, ~~\forall \bm x \in \partial \Omega_s\ \\
    \bm u\cdot \mathbf{n} = \bm v \cdot \mathbf{n}, ~~\textrm{on} \; \partial \Omega_f \cap \partial \Omega_s, 
    \label{eq:fluid_bc}
\end{dcases}
\end{equation}
where $\mathbf{n}$ denotes the surface normal of solid, and $\bm v$ is the solid velocity. The distance function $d(\cdot,\cdot)$ is defined as the shortest distance between an implicit interface and a point. 

On the free boundary, surface tension acts as a type of interfacial stress applied on the entire fluid surface $\partial \Omega_f$. For a liquid interface with uniform surface tension $\gamma$, the tangential stress is zero as there is no local surface tension gradient. The normal stress has to be balanced by the surface tension:
$
    \mathbf{n}^T\bm{\sigma}\mathbf{n}=\gamma(\nabla \cdot \mathbf{n}) 
$

The formulation above highlights two central challenges in solid–fluid coupling: enforcing non-penetration at the positional level and preserving fluid incompressibility in a consistent manner. Motivated by this observation, we seek a unified algorithmic framework that directly operates at the positional level and resolves solid and fluid states through a coupled optimization procedure while rigorously maintaining fluid volume. This perspective naturally leads to the prediction–optimization–correction scheme presented in the next section.

%% file: 4-numerical_algorithm.tex
\section{Numerical Algorithm}
The core idea of our method is based on the observation that the time integration schemes for both the solid and fluid components can be modified to follow a unified procedure consisting of three stages: \emph{position prediction}, \emph{energy minimization}, and \emph{velocity correction}, summarized as below:
\begin{enumerate}
\item Update velocities with external forces and predict positions;
\item Search for the final positions through energy minimization;
\item Compute the final velocities based on the final positions.
\end{enumerate}
This procedure enables a consistent treatment of incompressibility and non-penetration within a single variational framework. Concretely, the prediction stage advances each component independently under external forces; the optimization stage resolves elasticity, incompressibility, and contact constraints in a coupled manner at the positional level; and the correction stage recovers consistent velocities from the optimized configurations. This pipeline has been popular in simulating Lagrangian objects, where the energy minimization corresponds to the projection of physical constraints such as collision and contact \cite{muller2007position}. We generalize this idea to the Eulerian setting with fixed grid-based unknowns, enabling both components to share the same prediction–optimization–correction structure.

We present our algorithm by initially discussing the numerical schemes for the solid component (Section \ref{sec:solid}) and the fluid component (Section \ref{sec:fluid}) separately. In each section, we assume that the other component does not exist, allowing us to exemplify the \emph{prediction-optimization-correction} pipeline with concrete physical models, starting from the more intuitive solid component to the less intuitive fluid component. Next, we will introduce the key component of our algorithm, the coupling scheme along with its implementation details, in Section \ref{sec:coupl}.

\subsection{Lagrangian Solid}
\label{sec:solid}
Our solid component constitutes the discrete shells \cite{grinspun2003discrete} and rods \cite{bergou2008discrete}, discretized using triangles and segments separately. For shells, we incorporate membrane energy defined on triangles and hinge-based discrete bending energy defined on dihedral angles for the shell model. For rods, we include stretching energy defined on edges and bending energy defined on centerlines. 

The time integration of an elastic solid follows the \emph{prediction-optimization-correction} pipeline we presented above. 

First, we predict the intermediate position by employing external forces:
\begin{equation}
\bm x^{\star}=\bm x^n+\Delta t \bm v^n+\Delta t^2 \bm g.
\label{eq:predict_pos}
\end{equation}
Next, we solve the optimization problem of position to address elasticity \cite{gast2015optimization}:
\begin{equation}
\bm x^{n+1} = \mathop{\arg\min_{\bm x}}\frac{1}{2}\bm x^T \bm{M} \bm x-\bm x^T \bm{M} \bm x^{\star}+\Delta t^2 \Psi(\bm x),
\label{eq:solid_opt}
\end{equation}
where $\Psi(\bm x)$ is the elastic potential energy whose negative gradient is the interior elastic force. This formulation can be derived from implicit Euler time integration, where minimizing the incremental potential corresponds to enforcing the discrete momentum update. To interact with the environment, friction and contact can also be incorporated into Equation \ref{eq:solid_opt} as potential energies \cite{li2020incremental}.

Last, we conduct a velocity correction step to obtain the final velocity for the next time step: 
\begin{equation}
\bm v^{n+1}=(\bm x^{n+1}-\bm x^n)/\Delta t.
\label{eq:solid_vel_cor}
\end{equation}
For part of the examples, we also adopt the semi-implicit Rayleigh damping model described in \cite{chen2022unified} in the time integration. 

\subsection{Eulerian Fluid}
\label{sec:fluid}
Following the philosophy of Algorithm~\ref{alg:complete_scheme}, we devise a \emph{prediction-optimization-correction} scheme to solve the incompressible interfacial fluid dynamics. Our overarching goal is to design a numerical mechanism to rigorously enforce incompressibility on a positional level. 
We start from a classical advection-projection liquid solver (e.g., \cite{bridson2015fluid}) discretized on a staggered marker-and-cell (MAC) grid.
Scalar fields such as pressure $p$ are stored at cell centers, while vector fields such as velocity $\bm u$ are stored at face centers. We use a level-set field $\phi$ to track the fluid domain. Next, we will present the three steps, consisting of \emph{position prediction}, \emph{constrained optimization}, and \emph{velocity correction}, of the time integration scheme of a pure fluid system.

\subsubsection{Position Prediction}
First, we calculate the intermediate velocity field $\bm u^{\dagger}$ by applying external forces, including gravity and surface tension. Gravity is explicitly added to each face of the MAC grid. Surface tension is applied in a semi-implicit manner \cite{zheng2009simulation} by solving the following Poisson's equation on velocities within a narrow band $\varepsilon = 3\Delta x$ around the free surface, where $\Delta x$ is the grid cell size. These two steps can be combined as: 
\begin{equation}
(\bm{I}-\Delta t^2 \gamma \delta(\phi)\nabla^2)\bm u^{\dagger}=\bm u^n+\Delta t \bm g+\Delta t (-\gamma \delta(\phi)\kappa \mathbf{n}_f),\\
\label{eq:st}
\end{equation}
where $\gamma$ is the surface tension coefficient. $\delta(\phi)$ is a delta function defined within the narrowband to the surface as
$
\delta(\phi)=\frac{1}{2\varepsilon}(1+\cos(\frac{\phi \pi}{\varepsilon})).
$ $\mathbf{n}_f = \nabla \phi / \vert \nabla \phi\vert$ is the unit surface normal obtained from the level-set, and $\kappa = \nabla \cdot \mathbf{n}_f$ is the mean curvature. 

Next, we employ a projection step to obtain a divergence-free velocity field $\bm u^{\ddagger}$ from $\bm u^{\dagger}$ within the fluid domain.

Last, this divergence-free velocity field $\bm u^{\ddagger}$ is applied to predict the velocity and level set fields based on their values in the previous step as
\begin{equation}
(\bm u^{\star},\phi^{\star})=\textrm{Advection}(\bm u^{\ddagger}, \bm u^n, \phi^n, \Delta t).
\label{eq:advect_intermediate}
\end{equation}
In our implementation, we employed a standard semi-Lagrangian scheme for advection. We use linear interpolation and second order Runge Kutta as the tracing algorithm. Particle level-set is employed to track level-set accurately. The predicted level-set field $\phi^{\star}$ will be used as the predicted value for volume correction in Section~\ref{sec:fluid_opt} and the predicted velocity field $\bm u^{\star}$ will be used to obtain the final divergence-free velocity field at time $n+1$, as described in Section~\ref{fluid:vc}.

\subsubsection{Constrained Optimization}
\begin{figure}
 \centering
 \includegraphics[width=.45\textwidth]{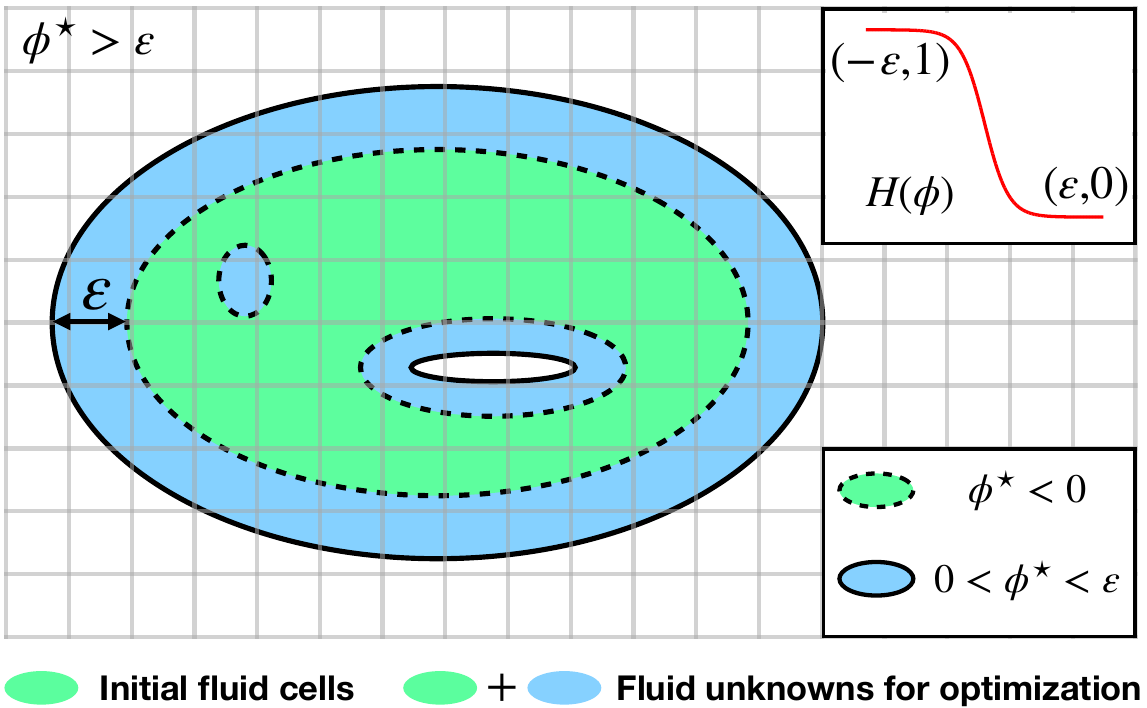}
 \caption{The level-set unknowns for optimization encompass grid cells within an $\varepsilon$-narrowband surrounding the initial interface. Including regions distant from the interface by more than $\varepsilon$ is unnecessary due to the nearly zero value of the Heaviside function in those areas, resulting in no impact on the fluid volume. The volume integral of the optimized level-set field is expected to precisely match the desired volume $V_0$.}
 \label{fig:phi_dof}
\end{figure}
\label{sec:fluid_opt}
After obtaining the predicted velocity and level set, we will solve a constrained optimization problem to rigorously enforce the incompressibility of the fluid volume on a positional level, by directly adjusting the level-set field near the interface. We formulate the optimization problem as motivated by the observation that the level-set field near its interface is smooth and differentiable, allowing the adjustment of a local fluid surface by editing its neighboring level-set values. Given an intermediate level-set domain $\phi^{\star}$ with its volume deviating from the original value $V_0$, we solve the following quadratic problem:
\begin{equation}
\begin{aligned}
    &\phi^{n+1} = \mathop{\arg\min_\phi}\frac{1}{2}\phi^T \bm{M}_f \phi-\phi^T \bm{M}_f \phi^{\star},\\
    &\bm{\mathrm{s.t.}} \;\; h(\phi)=\sum H(\phi)V_c-V_0= 0, 
\end{aligned}
\label{eq:phi_opt}
\end{equation}
where $\bm{M}_f=\rho V_c$ is a diagonal fluid mass matrix collected from the unknown cells. This objective corresponds to minimizing the weighted $L^2$ distance between $\phi$ and the predicted field $\phi^\star$, ensuring minimal modification to the intermediate interface while enforcing volume conservation. Here $V_c=\Delta x^d$ is the volume (or area in 2D) of a grid cell, and $\sum H(\phi)V_c$ is the discrete volume integral. For a level-set field with $n$ connected components, we calculate $V_0$ for each component and have $n$ corresponding volume constraints. Further, we define $H$ as a smoothed Heaviside step function:
\begin{equation}
\begin{dcases}
H(\phi)=\frac{1}{1+e^ {2k\phi}  },\\
H'(\phi)=-\frac{2ke^ {2k\phi}}{(1+e^ {2k\phi})^2},
\end{dcases}
\label{eq:heaviside}
\end{equation}
where $k$ controls the sharpness of the step function, which is set as the domain length $L$ divided by the narrowband width $\varepsilon$. To solve this problem using Newton's method $\phi^+=\phi+\Delta \phi$, it suffices to solve the following linear system for one Newton's step $\Delta \phi$:
\begin{equation}
\label{eq:newton_incomp}
\begin{bmatrix}
\bm H(\phi) & \bm J(\phi)^T\\
 \bm J(\phi) & \bm 0
\end{bmatrix}\begin{pmatrix}
\Delta \phi\\
\lambda\\

\end{pmatrix}=\begin{bmatrix}
-g(\phi) \\
-h(\phi)
\end{bmatrix},
\end{equation}
where $\bm H(\phi)$ and $g(\phi)$ are the Hessian matrix and the gradient vector of the objective function in Equation \ref{eq:phi_opt}. This linear system corresponds to the Karush–Kuhn–Tucker (KKT) conditions of the constrained optimization problem. $\bm J(\phi)$ is the Jacobian matrix of the constraint vector $h(\phi)$. The goal is to find a level-set configuration that is sufficiently close to the reference state $\phi^{\star}$ while satisfying the volume constraint.

We collect the group of unknowns $\phi$ for optimization based on the reference state $\phi^{\star}$. These unknowns include all grid cells distant from the initial state by less than a narrowband width $\varepsilon$, as illustrated in Figure \ref{fig:phi_dof}. The set of these cells remains unchanged during the Newton iterations. The selection of $\varepsilon$ depends on the CFL number, ensuring that the total fluid unknowns are sufficient to describe any deformation or topological change of the initial fluid during the optimization within the CFL-constrained time step size. Thus, including grid cells too far away from the initial state is redundant as they are guaranteed not to be affected during the optimization and contribute nothing to the fluid volume. Although the number of unknowns is fixed, the fluid mass matrix $\bm M_f$ is updated per iteration based on the current level-set values. The density of each grid cell follows a similar Heaviside distribution across the interface, changing smoothly from the liquid density $\rho_l$ ($-\varepsilon$) to the air density $\rho_a$ ($+\varepsilon$):
\begin{equation}
\rho =(\rho_l-\rho_a)\frac{1}{1+e^ {2k\phi}  }+\rho_a.
\label{eq:heavi_density}
\end{equation}

For the initial guess, we use $\phi^n$ from the previous time step. A redistancing scheme (fast marching) is only applied once after the optimization, which is not employed during the Newton iterations even though the level-set values are changing. Another concern would be what if the fluid interface crosses the border of the narrowband during the optimization. Since the maximum displacement of the fluid within a time step is restricted by the CFL condition, we use $\varepsilon=3 \times \mathrm{MaxDisplacement}$ for safety such that the narrowband level-set is always sufficient to describe the interface throughout the optimization.

It is worth noting that all components in the above linear system manifest very simple forms and the optimization step is inexpensive to perform. We briefly explain these computational merits in several aspects.
First, the objective is quadratic and converges fast with Newton-type solvers. Second, the Hessian matrix part in Equation \ref{eq:newton_incomp} is diagonal and can be factorized with ease. Lastly, since we already obtained a divergence-free velocity field from projection, the advected level-set is not far from satisfying the volume constraint, so the adjustment in optimization is moderate. Thus, both the external Newton solver and the internal sparse solver require very minimum steps (typically less than 5 steps) to converge.

\subsubsection{Velocity Correction}
\label{fluid:vc}
After obtaining the corrected level set, we will conduct another projection step to obtain the final divergence-free velocity field on the corrected level-set domain. We summarize the full time integration scheme for a pure fluid system in Algorithm \ref{alg:fluid_opt}. Note that an extrapolation process to propagate fluid velocities into the surrounding air is employed by default after each projection step, which is omitted in the algorithm.

\begin{algorithm}[t]
\caption{Fluid Time Integration Scheme.}
\label{alg:fluid_opt}
\begin{flushleft}
\algorithmicrequire{ $\bm u^{n}$, $\phi^{n}$, $\bm g$, $\gamma$, $\rho$, $V_0$, $\Delta t$} \hspace{0.3 in} \algorithmicensure{ $\bm u^{n+1}$, $\phi^{n+1}$}
\end{flushleft}
\begin{algorithmic}[1]
\State $\bm u^{\dagger}\leftarrow$ApplyForces($\bm u^{n}$, $\phi^{n}$, $\bm g$, $\gamma$, $\Delta t$) \Comment{Eq. \ref{eq:st}}
\State $\bm u^{\ddagger}\leftarrow$Projection($\bm u^{\dagger}$, $\phi^{n}$, $\Delta t$) \Comment{Poisson}
\State $\bm u^{\star}\leftarrow$Advection$^{\bm u}$($\bm u^{\ddagger}$, $\bm u^{n}$, $\Delta t$) \Comment{SemiLagrangian}
\State $\phi^{\star}\leftarrow$Advection$^{\phi}$($\bm u^{\ddagger}$, $\phi^{n}$, $\Delta t$) \Comment{SemiLagrangian}
\State $\phi^{n+1}\leftarrow$Optimization$^{\phi}$($\phi^{\star}$, $\phi^n$, $\rho$, $V_0$) \Comment{Eq. \ref{eq:phi_opt}, \ref{eq:heaviside}, \ref{eq:newton_incomp}}
\State $\bm u^{n+1}\leftarrow$Projection($\bm u^{\star}$, $\phi^{n+1}$, $\Delta t$) \Comment{Poisson}
\end{algorithmic}
\textbf{Note:} $\bm{u}^\dagger$ is the intermediate velocity field with external force applied. $\bm{u}^\ddagger$ is the divergence-free velocity field projected from $\bm{u}^\dagger$. $\bm{u}^\star$ is the predicted velocity field.
\end{algorithm}
\begin{figure}
 \centering
 \includegraphics[width=.3\textwidth]{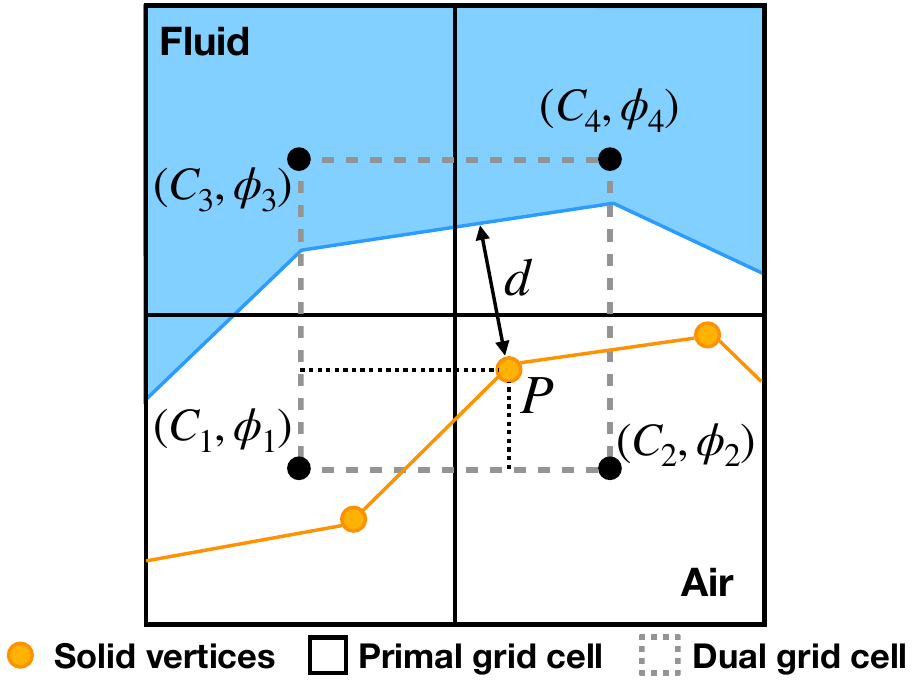}
 \caption{A primitive pair between the Eulerian fluid and the Lagrangian solid comprises a vertex and the corresponding dual grid cell it occupies. The distance between them is determined by a weighted linear combination of the level-set values at the adjacent primal grid cells, taking into account their relative positions.} 
 \label{fig:interp}
\end{figure}
\subsection{Contact Handling}

\label{sec:coupl}
Having explained the individual frameworks for the solid and fluid components, we now proceed to discuss their coupling. We introduce a novel coupling mechanism that defines and discretizes the necessary constraints to handle the contact between the two components. As mentioned earlier in Equation~\ref{eq:fluid_bc}, two essential constraints ensure a robust solid-fluid contact: (1) avoidance of penetration between the solid and fluid volumes, and (2) consistency of velocities in the contact regions. We will demonstrate how these constraints can be formulated in their forms and effectively solved using optimization solvers.

\subsubsection{Distance Metric}
Given a point $\bm x$, the level-set field automatically gives the signed distance from the query point to the closest point on the interface:
\begin{equation}
d(\phi=0, \bm x)=\phi(\bm x).
\label{eq:phi_intp}
\end{equation}
Without sacrificing clarity, we use $d(\phi,\bm x)$ to replace $d(\phi=0,\bm x)$ for conciseness. In the discrete world, this signed distance can be approximated via interpolation on the background grid. Taking the simplest case (2D bilinear interpolation) for example, point $\bm P$ stays in a dual grid cell as shown in Figure \ref{fig:interp}, whose four incident nodes correspond to the centers of the primal grid cells $\bm C_1-\bm C_4$. The distance is calculated as:
\begin{equation}
\begin{aligned}
d(\phi, \bm x)=& \phi_1(1-\bm f_x)(1-\bm f_y)\\
&+\phi_2\bm f_x(1-\bm f_y)\\
&+\phi_3(1-\bm f_x)\bm f_y\\
&+\phi_4\bm f_x \bm f_y,
\label{eq:dist}
\end{aligned}
\end{equation}
where $\bm f$ is the relative position fraction with respect to the bottom left cell $\bm C_1$:
\begin{equation}
\label{eq:frac}
\bm f=\begin{bmatrix}
\frac{\bm P_x-\bm C_{1x}}{\bm C_{2x}-\bm C_{1x}} \\[2ex]
\frac{\bm P_y-\bm C_{1y}}{\bm C_{3y}-\bm C_{1y}}
\end{bmatrix}=\begin{bmatrix}
\frac{\bm P_x-\bm C_{1x}}{\Delta x} \\[2ex]
\frac{\bm P_y-\bm C_{1y}}{\Delta x}
\end{bmatrix}.
\end{equation}

\begin{figure}
 \centering
 \includegraphics[width=.3\textwidth]{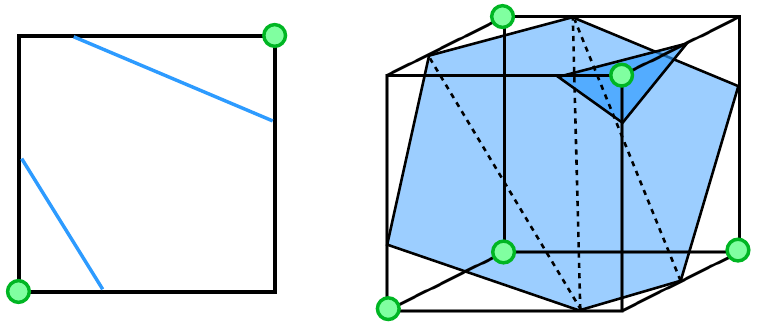}
 \caption{Sample potential geometries depicted by the nodal level-set values on dual cells. Green dots indicate points located inside the fluid ($\phi<0$).} 
 \label{fig:patch}
\end{figure}

An intuitive way to understand this distance metric is that each dual cell represents a virtual "patch" of the fluid surface, which is then paired with neighboring solid vertex candidates. This small "patch" has better versatility and expressiveness than the basic simplicial complex elements such as edge and triangle. Parametrized by 4 level-set values (8 in 3D), the "patch" can represent complex geometries, and even disjoint segments and planes, as shown in Figure \ref{fig:patch}. Equation \ref{eq:dist} naturally extends to 3D as trilinear interpolation. High-order interpolation schemes, such as quadratic spline interpolation can match better to the local fluid surface by leveraging more level-set samples from neighboring cells, providing more accurate distance results. However, higher-order methods place stronger demands on the smoothness of the sampling data. At sharp corners of the fluid, where the level-set is not smooth enough, overly high-order interpolations can lead to severe oscillation errors, harming the convergence of the optimization solver. We thus limit our implementation to linear and quadratic interpolation as a compromise of efficiency and accuracy.

\subsubsection{Barrier Formulation and Differentiation}
With the distance metric between the solid and the fluid surface defined above, the penetration-free constraint can be described by the following inequality:
\begin{equation}
d(\phi,\bm x)\geq 0,
\label{eq:dist_const}
\end{equation}
for any point $\bm x$ on the solid boundary. Inspired by \cite{li2020incremental}, the above constraint is equivalent to incorporating an additional barrier energy term to the objective:
\begin{equation}
E_b(\phi, \bm x)=\varkappa \sum_c b(d(\phi, \bm x)),
\label{eq:barrier_ene}
\end{equation}
where $\varkappa$ is a barrier stiffness; $c$ is the primitive pair group; $b$ is a $C^2$-smooth barrier function which provides arbitrarily large repulsion to prevent the distance metric from being non-positive. In particular, the expression of $b$ can be defined as: 
\begin{equation} 
b(d,\hat{d})=
\begin{dcases}
    -(\frac{d}{\hat{d}}-1)^2\ln{\frac{d}{\hat{d}}}, ~~0<d<\hat{d},\\
    0,~~d\geq \hat{d},
    \end{dcases}
\label{eq:barrier_func}
\end{equation}
where $\hat{d}$ is a distance clamping threshold such that the active primitive pair group $c$ only contains those closer than the threshold. Each primitive pair consists of a solid vertex and its associated dual cell (see Figure \ref{fig:interp}). 
The first and second derivatives of the barrier function $\frac{\partial b}{\partial d}$ and $\frac{\partial^2 b}{\partial d^2}$ are easy to find and have $C^1$, $C^0$-smoothness respectively.

Optimizing the barrier energy proposed in Equation \ref{eq:barrier_ene} requires computing its gradient and Hessian:
\begin{equation}
\begin{dcases}
\nabla b=\frac{\partial b}{\partial d}\frac{\partial d}{\partial (\phi,\bm x)},\\
\nabla^2 b\approx\frac{\partial d}{\partial (\phi, \bm x)}^T\frac{\partial ^2 b}{\partial d^2}\frac{\partial d}{\partial (\phi, \bm x)},
\end{dcases}
\label{eq:barrier_gh}
\end{equation}
We omit $\frac{\partial b}{\partial d}\frac{\partial^2 d}{\partial(\phi, \bm x)^2}$ in the Hessian because this term is numerically small. Because the signed distance $d(\phi,\mathbf{x})$ is obtained from linear or quadratic interpolation, its derivative $\partial d / \partial(\phi,\mathbf{x})$ involves only low-order polynomial basis functions, whose gradients vary smoothly. In contrast, the gradient and curvature of the barrier function $b$ change much more rapidly. Consequently, the contribution of the mixed term is negligible compared to the dominant Hessian term.

For linearly interpolated distance (\ref{eq:dist}), the gradient is:
\begin{equation}
\label{eq:dist_grad}
\frac{\partial d}{\partial (\phi, \bm x)}=\begin{bmatrix}
\frac{\partial d}{\partial \phi_{1-4}} \\[2ex]
\frac{\partial d}{\partial \bm x_{x,y}}
\end{bmatrix}=\begin{bmatrix}
(1-\bm f_x)(1-\bm f_y) \\[1ex]
\bm f_x(1-\bm f_y) \\[1ex]
(1-\bm f_x)\bm f_y \\[1ex]
\bm f_x\bm f_y \\[1ex]
\frac{(\phi_2-\phi_1)(1-\bm f_y)+(\phi_4-\phi_3)\bm f_y}{\Delta x}\\[1ex]
\frac{(\phi_3-\phi_1)(1-\bm f_x)+(\phi_4-\phi_2)\bm f_x}{\Delta x}
\end{bmatrix},
\end{equation}
where $\frac{\partial d}{\partial (\phi, \bm x)}$ is the concatenation of $\frac{\partial d}{\partial \phi_{1-4}}$ and $\frac{\partial d}{\partial \bm x_{x,y}}$. $\frac{\partial d}{\partial \phi_{1-4}}$ is the partial derivative of the signed distance $d$ with respect to the level-set values on the centers of the primal grid cells $\bm C_1-\bm C_4$. Each barrier Hessian is constructed as a small $6\times 6$ matrix ($11\times 11$ in 3D) restricted to the vertex and its associated level-set data. Gradient and Hessian forms for quadratic interpolation can be derived in a similar manner.

\subsubsection{Global Smoothness and Guaranteed Collision-Free}
\label{sec:intp_scheme}

\begin{figure}
 \centering
 \includegraphics[width=.4\textwidth]{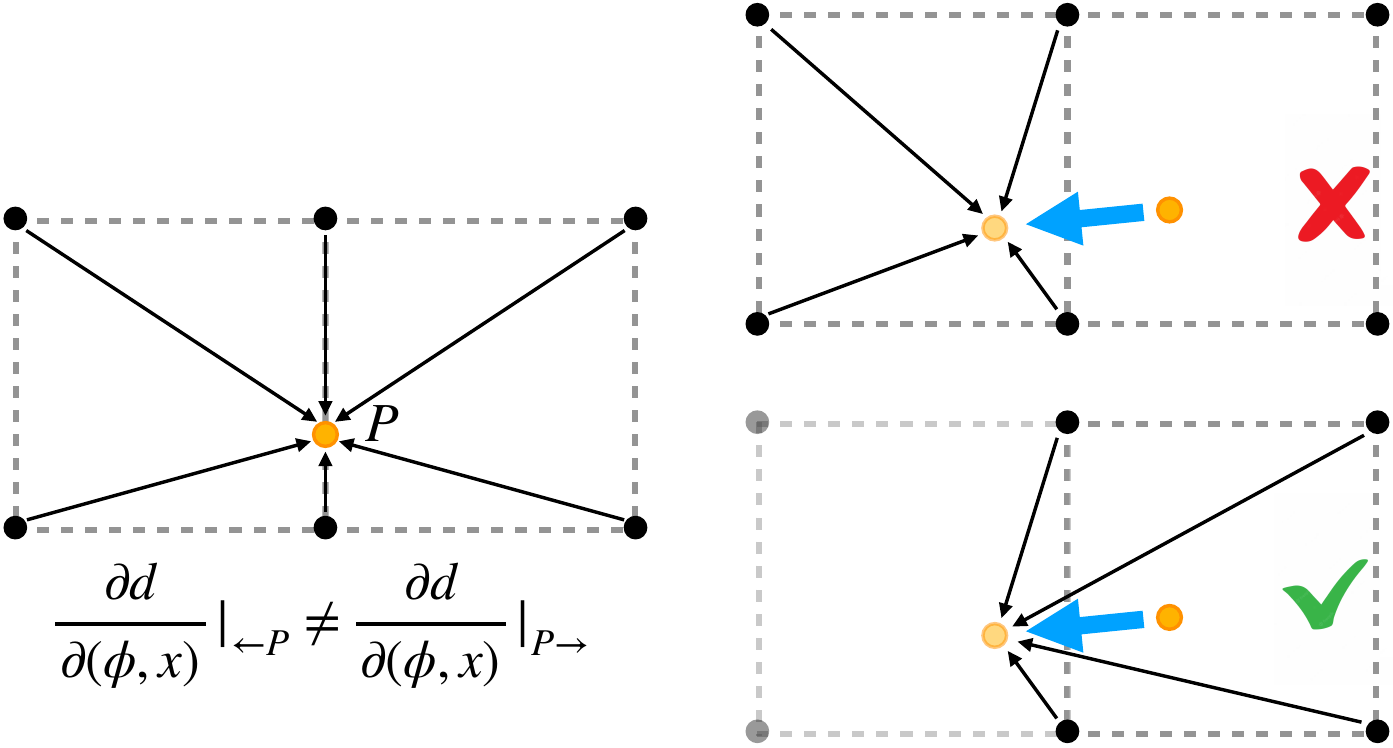}
 \caption{Left: Gradient ambiguity arises due to linear interpolation at the boundaries of dual cells. Right: When a solid vertex moves beyond its original dual cell during the Newton iteration, the originally associated level-set samples are still used to estimate the current distance to the fluid surface.} 
 \label{fig:smooth}
\end{figure}

While embracing the simplicity of linear interpolation provided by Equation \ref{eq:dist}, a major drawback is the nonsmoothness of its gradient (Equation \ref{eq:dist_grad}) across the dual cell's border, which can significantly degrade the convergence of Newton-type solvers. As shown in Figure \ref{fig:smooth} left, although the interpolated signed distance at $P$ is unique, the distance gradient is discontinuous on each side and involves different groups of unknowns. During the Newton iterations, solid vertices can always move between different dual cells, thus making the system nonsmooth. To make Equation \ref{eq:dist_grad} also continuous, we collect the primitive pair group $c$ before optimization based on the reference state $(\phi^{\star}, \bm x^{\star})$, and keep the spatial relations between solid vertices and dual cells unchanged during the Newton iterations. As shown in Figure \ref{fig:smooth} right, when a solid vertex moves from its originally paired dual cell to another, instead of interpolating the distance based on the current cells, we keep using values from the old cells for extrapolation. As a result, the distance evaluation can be inaccurate, particularly when the time step size is large and solid vertices are away from their originally paired cells.

For higher-order interpolation schemes that satisfy at least $C^1$ smoothness everywhere, such issue no longer exists and the primitive pair group $c$ can be updated at each Newton iteration based on the solid vertices' current locations, which is implemented along with our quadratic interpolation scheme. Since the spatial relations are updated, the distances are always interpolated and the accuracy is no longer relevant to the time step size. The cost is a more complex gradient form and a larger local Hessian. A detailed validation of both schemes' performances is provided in Section~\ref{sec:val}.

We also want to highlight that point-dual-cell is the only type of primitive pair we use for different solid geometries. The barrier energy can thus ensure the level-set values for all solid vertices are positive. This treatment is weak as penetrations can happen between fluids and codimension-1 triangles and codimension-2 segments. However, the level-set method inherently cannot represent sharp features on the subcell level ($\kappa < \frac{1}{\Delta x}$). As long as the lengths of all codimension-2 edges are less than the grid cell size $\Delta x$ during the simulation, it can be expected that no penetration will happen between fluids and all types of codimensional elements. Note that the triangle width is also constrained by its longest edge. 

\subsubsection{Optimization}
\label{sec:optimization}

We solve the following optimization problem by combing all the three objectives specified in Equation~\ref{eq:solid_opt}, ~\ref{eq:phi_opt}, and ~\ref{eq:barrier_ene}:
\begin{equation}
\begin{aligned}
    (\phi^{n+1},\bm x^{n+1}) = 
    &\mathop{\arg\min_{\phi,\bm x}}\underbrace{\frac{1}{2}\phi^T \bm{M}_f \phi-\phi^T \bm{M}_f \phi^{\star}}_{\text{\emph{fluid inertia}}}\\
    &+\underbrace{\frac{1}{2}\bm x^T \bm{M} \bm x-\bm x^T \bm{M} \bm x^{\star}+\Delta t^2 \Psi(\bm x)}_{\text{\emph{solid inertia, elastic potential and damping}}}\\
    &+\underbrace{\varkappa \sum_c b(d(\phi, \bm x))}_{\text{\emph{contact barrier}}},\\
    &\bm{\mathrm{s.t.}} \;\; h(\phi)=\underbrace{\sum H(\phi)V_c-V_0= 0}_{\text{\emph{incompressibility}}}.
\end{aligned}
\label{eq:complete_opt}
\end{equation}
Note that the optimization steps search for the final configuration that satisfies collision-free while maintaining the fluid volume. We do not explicitly model the momentum exchange through the coupling, which is implicitly resolved by the optimizer using the relative magnitude of the fluid mass, solid mass and solid material property in the combined Hessian matrix, as well as the constraints on incompressibility and collision-free. 

Following \cite{li2020incremental}, we use a Newton-type solver to find the new configuration. The Hessian parts of the fluid (Equation \ref{eq:newton_incomp}) and the solid are pre-allocated and assembled separately, which are then merged together. The barrier Hessian does not need to be explicitly built, which is instead analyzed locally for each primitive pair and added to the corresponding entries of the complete Hessian. For termination we check whether the $\ell^{2}$-norm of the Newton's direction scaled by the time step size is less than a threshold (e.g. 4e-2).

The Newton method is initialized using the configuration from the previous step, where there is no interpenetration. The independently predicted solid and fluid positions are allowed to penetrate. The optimizer then searches for an intermediate state that satisfies all constraints based on the relative mass and material properties.
\begin{figure}
 \centering
 \includegraphics[width=.45\textwidth]{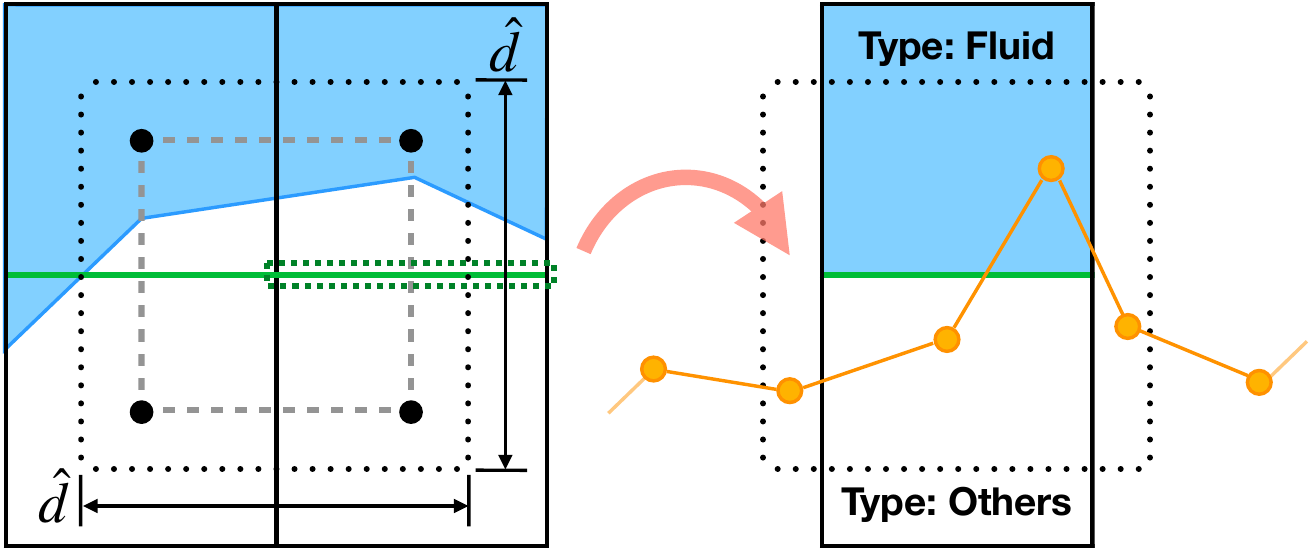}
 \caption{Left: Fluid-air interfaces (depicted in green) associated with the dual cell in contact with the solid. Right: The facial velocity is determined by the nodal velocities of the solid nodes incident to it.} 
 \label{fig:solid_bc}
\end{figure}

\subsubsection{Convergence of Newton}
\label{sec:convergence}
Newton's method is only locally convergent, and quadratic convergence of Newton takes effect only when we start sufficiently close to a root. While Eulerian-based fluid simulations typically require small time step sizes for robustness, there's no guarantee that the initial configuration already resides in the quadratic zone. Another notable concern is that the optimization process can violate the interpenetration-free constraint, even if the Newton solve is converging. To address these issues, we design a set of strategies to first, ensure that the Newton iterations genuinely converge to a minimum, and second, prevent interpenetration throughout the optimization.

The convergence of Newton requires the Newton direction to be a descent direction, with a sufficient condition being that the Hessian is positive definite. Following \cite{chen2022simulation}, we approximate our Hessian to be positive definite by adding a regularizer. To ensure the residual norm is always decreasing, we employ backtracking line search \cite{jorge2006numerical} to avoid overshooting. To confirm quadratic convergence, we check the Newton step size at the final iteration to see if it is close to 1.

To prevent interpenetration throughout the Newton iterations, we devise a continuous collision detection (CCD) filter to control the upper bound of the Newton step size. As shown in Algorithm~\ref{alg:line_search_ccd}, we estimate the time of impact (TOI) of a primitive pair as follows: Suppose a solid vertex $i$ moves from $x_i$ to $x_{i,p}=x_i+p_i$ with $p$ being the Newton direction, we find the corresponding closest points on the fluid surface before and after the Newton direction by projecting along the normal provided by the level-set field, denoted as $x_{\phi}$ and $x_{\phi,p}$. This gives us two segments: $S_p: x_i \rightarrow x_{i,p}$ and $S_{\phi}: x_{\phi} \rightarrow x_{\phi,p}$. A reasonable TOI can then be estimated by finding $t\in[0,1]$ that minimize the distance between the two segments, $d=\vert\vert S_p(t)-S_{\phi}(t) \vert\vert$. To further simplify the computation, we use a lower bound of this TOI by finding the shortest distance between the two segments. Solving a 2 by 2 linear system provides $\gamma, \beta \in[0,1]$ such that $S_p(\gamma)\rightarrow S_{\phi}(\beta)$ forms the shortest path between $S_p$ and $S_{\phi}$. It's clear that $t>\min(\gamma, \beta)$, and we employ $t_b=\min(\gamma, \beta)$ as the upper bound of the Newton step size used in the line search. For efficiency concerns, this collision detection filter is only applied to primitive pairs whose distances evaluate as negative after a complete Newton step.

Line search with CCD is an expensive process, particularly prominent under small time steps. In the case of Eulerian-based fluids, a strong CFL restriction is inevitable for stability, especially under large surface tension. Through experiments, we find that for very small time step sizes, dropping the line search won't violate the constraints or affect the convergence. A detailed analysis on line search and convergence of Newton is provided in Section 5.1.

\begin{algorithm}[t]
\caption{Line Search with CCD}
\label{alg:line_search_ccd}
\begin{flushleft}
\algorithmicrequire{ $x_i, p_i, \phi$} \hspace{0.3 in} \algorithmicensure{ $t_b$}
\end{flushleft}
\begin{algorithmic}[1]
\State $x_{i, p}\leftarrow x_p+p_i$
\State $x_\phi\leftarrow\text{Project}(x_i, \phi), x_{\phi,p}\leftarrow\text{Project}(x_{i, p}, \phi)$
\State $S_p:x_i\rightarrow x_{i,p}$, $S_\phi: x_\phi\rightarrow x_{\phi,p}$
\State $\gamma, \beta\leftarrow\arg\min_{t_1, t_2\in[0,1]}|S_p(t_1)-S_\phi(t_2)|$
\State $t_b\leftarrow\min(\gamma, \beta)$
\end{algorithmic}
\end{algorithm}

\subsubsection{Velocity Correction with Contact}
Lastly, we apply a velocity correction step to ensure the solid and fluid components have consistent velocities in contact regions after solving the constrained optimization problem. As shown in Algorithm~\ref{alg:complete_scheme} Line 9 and Equation \ref{eq:solid_vel_cor}, the velocity correction for solid is to trivially calculate each vertex's velocity with the corrected position. For fluid velocity, we conduct a standard projection step with additional Neumann boundary conditions served by the contacting solids, as shown in Line 10 and 11 in Algorithm \ref{alg:complete_scheme}. We first collect the set of dual cells in the primitive pair group $c$ after the optimization, which are considered in contact with the solid (distant by less than $\hat{d}$). For each dual cell, we search for grid faces sandwiched by different types of cells within a range of $\hat{d}$, as shown in Figure \ref{fig:solid_bc} left. Next, we collect all solid elements that have intersections with a $\hat{d}$-sized box centered at the face, and the facial velocity is a weighted sum of the velocities from the associated solid vertices, as shown in Figure \ref{fig:solid_bc} right:
\begin{equation}
\bm u^{n+1}=\bm W \bm v^{n+1}.\\
    \label{eq:u_n+1}
\end{equation}
The weight of each vertex is proportional to the sum of its incident elements' overlapping areas with the box. Each row of the weight matrix $\bm W$ is normalized to a unit. A detailed explanation for $\bm W$ is provided by \cite{robinson2008two}.
After completing the velocity correction step, our final configurations of $(\mathbf{u}^{n+1}, \phi^{n+1}, \mathbf{v}^{n+1}, \mathbf{x}^{n+1})$ satisfy both interpenetration-free constraint in terms of position and coherence in terms of velocity.

\begin{algorithm}[t]
\caption{Coupling System Time Integration Scheme.}
\label{alg:complete_scheme}
\begin{flushleft}
\algorithmicrequire{ $\bm u^{n}$, $\phi^{n}$, $\bm v^{n}$, $\bm x^n$} \hspace{0.3 in} \algorithmicensure{ $\bm u^{n+1}$, $\phi^{n+1}$, $\bm v^{n+1}$, $\bm x^{n+1}$}
\end{flushleft}
\begin{algorithmic}[1]
\State \textbf{Position Prediction}
\State $\quad$ \textbf{S:}$\; \bm x^{\star}\leftarrow$Advance($\bm v^{n}$, $\bm x^{n}$, $\Delta t$) \Comment{Eq. \ref{eq:predict_pos}}
\State $\quad$ \textbf{F:}$\;\phi^{\star}, \bm u^* \leftarrow$Advance($\bm u^{n}$, $\phi^{n}$, $\Delta t$) \Comment{Alg. \ref{alg:fluid_opt}: 1-4}
\State 
\State \textbf{Constrained Optimization} 
\State $\quad$ $\bm x^{n+1}, \phi^{n+1}\leftarrow$Optimization$^{\bm x, \phi}$($\bm x^{\star}$, $\bm x^{n}$, $\phi^{\star}$, $\phi^n$, $\Delta t$) 
\State \Comment{Eq. \ref{eq:solid_opt}, \ref{eq:phi_opt}, \ref{eq:barrier_ene}}
\State \textbf{Velocity Correction}
\State $\quad$ \textbf{S:}$\; \bm v^{n+1}\leftarrow$($\bm x^{n+1}$, $\bm x^{n}$, $\Delta t$) \Comment{Eq. \ref{eq:solid_vel_cor}}
\State $\quad$ \textbf{F:}$\; \bm u^{n+1}\leftarrow$Projection($\bm u^{\star}$, $\phi^{n+1}$, $\bm v^{n+1}$, $\bm x^{n+1}$, $\Delta t$) 
\State $\qquad$ $\, \bm u^{n+1}\cdot \mathbf{n}=\bm  v^{n+1} \cdot \mathbf{n}, \forall d(\phi^{n+1}=0, \bm x^{n+1})<\hat{d}$ \Comment{Poisson}
\end{algorithmic}
\end{algorithm}

To summarize, as outlined in Algorithm~\ref{alg:complete_scheme},  the solid and fluid states are first independently predicted through time integration. Next, a coupled constrained optimization jointly updates solid positions and the fluid level set by enforcing non-penetration and incompressibility at the positional level. Finally, velocities are corrected from the optimized configurations, with a projection step ensuring normal velocity consistency in contact regions.

%% file: 5-results.tex
\section{Results}

We first conducted validation experiments in 2D to provide intuitive visualization, with a focus on assessing the efficacy and accuracy of our major contributions. Subsequently, we showcase simulations at various scales to emphasize the superiorities of our approach in contact handling, energy preservation, and overall versatility. Towards the end, we include performance analyses by breaking down the time consumption of a single step.

\subsection{Validation}
\label{sec:val}

\begin{figure}
\centering
 \includegraphics[width=.44\textwidth]{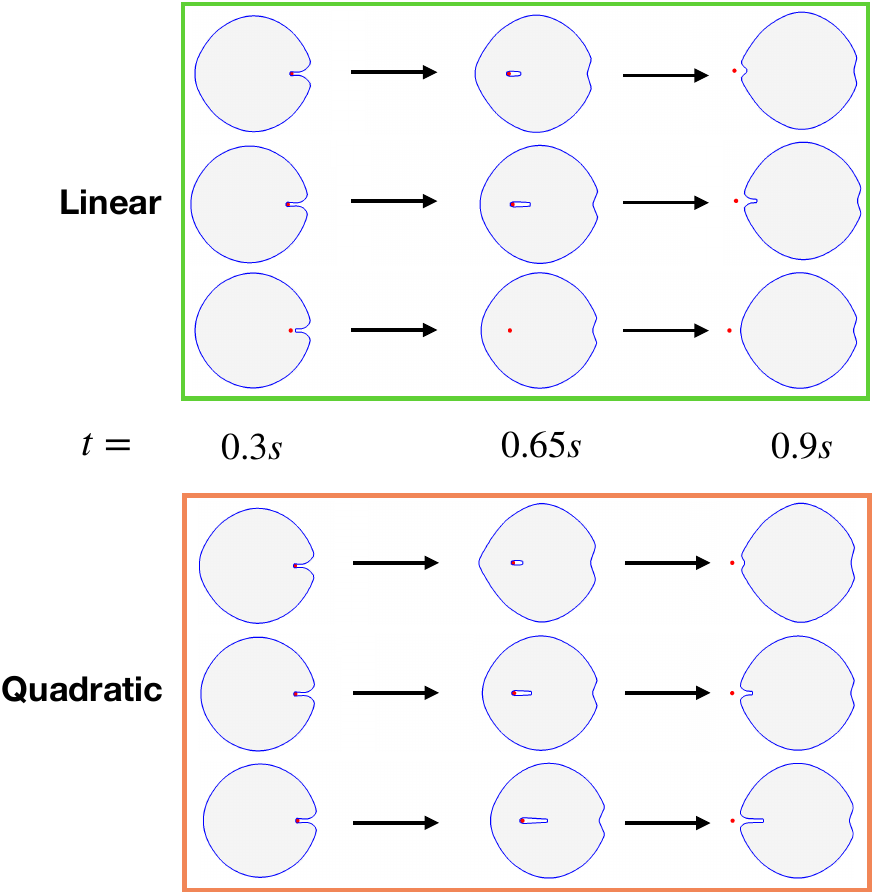}
 \caption{A single solid vertex(red) collides with a fluid (pale) at high speed under zero gravity. Both interpolation schemes are tested under three different CFL conditions: 0.5 (top), 1.0 (mid), 3.0 (bot). Frames at $t=0.3s, 0.65s, 0.9s$ are selected to show the topological changes of the fluid.}
 \label{fig:val_intp}
\end{figure}

\begin{figure*}
\centering
 \includegraphics[width=1.\textwidth]{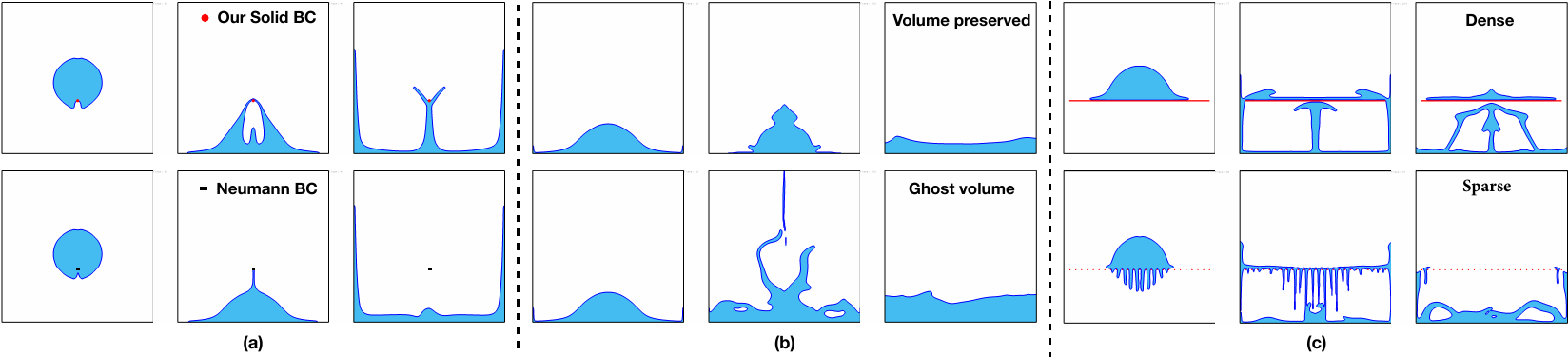}
 \caption{A series of 2D validation tests. (a): Comparison of collision handling results between a fluid volume and a single fixed point using our method (top) and Neumann boundary condition (bottom). (b): Comparison of volume change when a droplet impacts the ground at high speed with our incompressibility constraint technique turned on (top) and off (bottom). (c): Study of liquid permeability based on solid porosity. The solid vertices have a uniform spacing of $1\times$ cell size (top) and $5\times$ cell size (bottom).}
 \label{fig:validation}
\end{figure*}

\textbf{Interpolation scheme.}
As discussed in Section~\ref{sec:intp_scheme}, two interpolation schemes (linear and quadratic) are employed in our framework to estimate the distances between solids and fluids. While linear interpolation offers simpler gradient and Hessian forms, the nonsmoothness of the derivative necessitates the use of the initial primitive pairs throughout the optimization process. This introduces significant inaccuracies, especially at large time step sizes. In contrast, quadratic interpolation is inherently smooth and avoids such issue. In Figure \ref{fig:val_intp}, we let a droplet and a particle collide with each other. To ensure interpenetration-free interactions, a small air bubble should form around the particle as it passes through the droplet. The bubble exists because it takes some time for surface tension to close the gap. We conduct this experiment for both interpolation schemes under various CFL conditions. Upon observation, quadratic interpolation behaves as expected across a range of CFL numbers, from small to large. Linear interpolation behaves similarly to quadratic interpolation at small to moderate CFL numbers, but fails at large CFL numbers and causes interpenetration.

\textbf{Contact handling.} 
The simplest and most convincing way to highlight the contact handling capacity of our approach is to collide a liquid volume by a single particle, which is more visible in 2D. In Figure \ref{fig:validation} (a), we put a static particle in space (marked in red) whose mass is much larger than the liquid such that it serves similarly to a rigid boundary of infinitesimal size. We then pour a droplet on it with an initial speed of $-1m/s$ in the $y$-direction. In comparison, we directly add a Neumann boundary condition at the same location of the particle (marked in black) to enforce the fluid velocity to be zero on that face. The results show that our position-level penetration-free constraint correctly handles the collisions from all directions even for a single dimensionless particle. On the other hand, enforcing a velocity-level constraint fails to prevent the solid boundary from penetrating the liquid. An additional sticky artifact is also observed when the droplet passes through the obstacle.
\begin{figure}[t]
\includegraphics[width=0.48\textwidth]{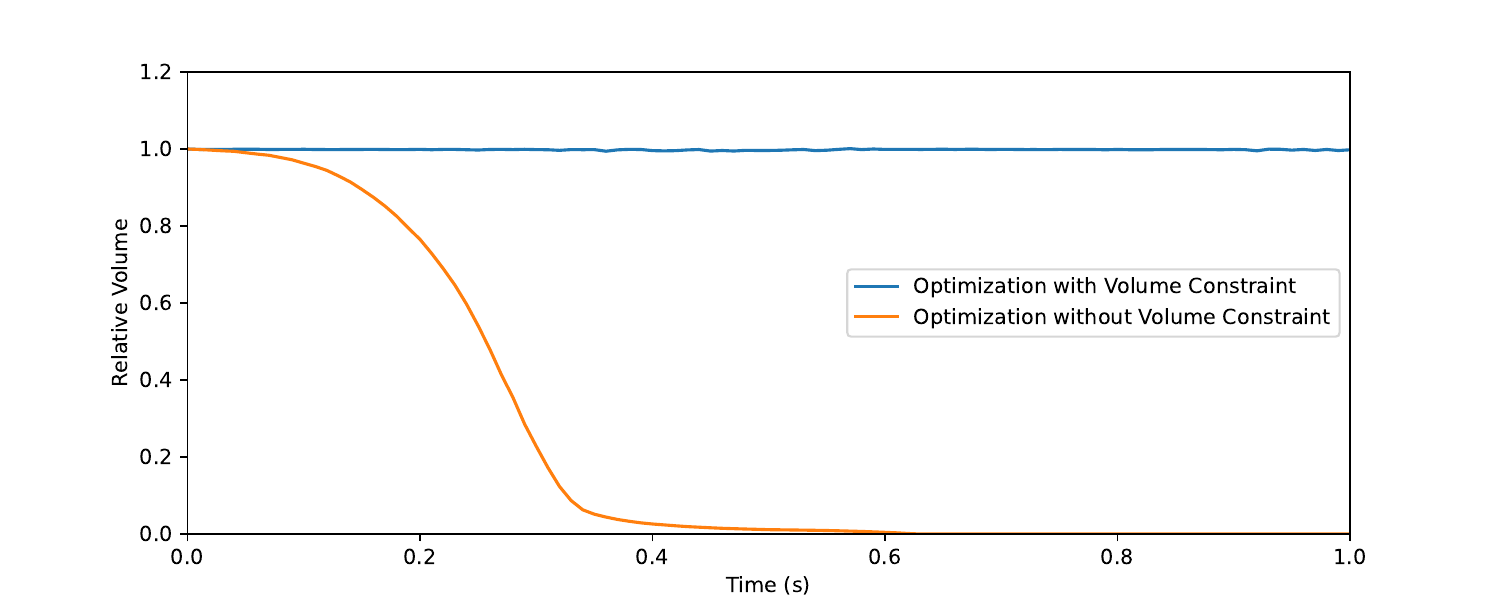}
\caption{Relative volume of the level-set in the example Bouncing. Results of optimization with/without volume constraint are plotted.}
\label{fig:volume}
\end{figure}

\textbf{Fluid incompressibility.}
Volume conservation of the level-set is critical for accurately capturing fluid details and multiphase contacts. Compared to existing approaches, our method is simple in theory, straightforward to implement, cheap in computational cost yet still effective, which can work as a standalone module or couple with other parts with ease. In Figure \ref{fig:validation} (b), we let a droplet hit the ground at a high speed of $-5m/s$, causing the liquid to climb along side walls. Without our incompressibility module (bottom row), we see drastic volume changes, especially when the liquid parts from top and bottom collide and merge causing a lot of air voids suddenly disappear. This issue is mainly caused by the unprecise level-set representation at sharp corners and erroneous advection schemes. On the contrary, as shown in the top row, our method corrects unphysical volume gain or loss within each time step. The complete simulation is stable without any accumulated volume error. For the 3D case, we report the relative volume of the level set in the Bouncing example (Figure \ref{fig:volume}). As an ablation, we also show the result of the optimization without volume constraint, where the volume quickly collapses. In contrast, our optimization with volume constraint maintains the volume well, with error below 1\%.

\begin{figure}
 \centering
 \includegraphics[width=.35\textwidth]{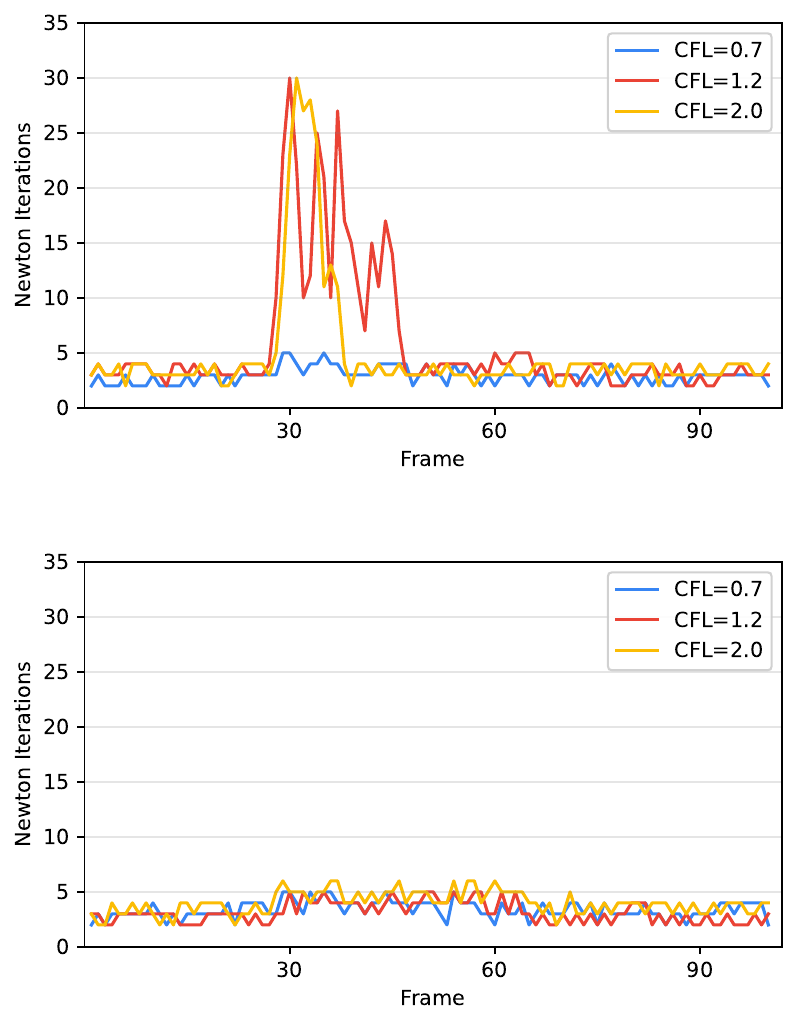}
 \caption{Maximum number of Newton iterations at each frame under different CFL conditions with (bottom) and without (top) employing backtracking line search with CCD.}
 \label{fig:ls_conv}
\end{figure}

\textbf{Particle spacing and penetration.}
As mentioned before, our contact handling between solids and fluids is based on particle, and no edge-fluid or triangle-fluid distance metric has been defined. Thus, collision-free relies on a relatively dense particle distribution compared to the grid with a cell size of $\Delta x$, such that the level-set "sees" the discrete particles as continuous segments or surfaces. The question arises as what is the largest spacing between particles before penetration happens. An accurate quantification of the extreme spacing is impossible, because penetration also depends on many other factors, such as the distance threshold and solving accuracy customized by users. However, we do observe a continuous penetration behavior by gradually decreasing the solid's resolution. An example is shown in Figure \ref{fig:validation} (c), where a droplet whose initial speed is $(0,-1m/s)$ falls on a row of fixed particles. For the top row, the distance between adjacent particles is equal to the grid cell size, and we observe a smooth contact behavior as if the particles form a tight thin wall. For the bottom row, the spacing is increased by 5 times, and we see penetrations between particle gaps as if passing through a leaky wall. Meanwhile, penetration-free for each single particle is still secured. For this test, we find that jiggering and bumpy contacts start when the particle spacing is at around $1.8\Delta x$. We also experiment extreme conditions by increasing the initial falling speed and reducing the distance threshold, and conclude that a particle spacing of $1.2\Delta x$ or below is generally safe enough to avoid any penetration.

\textbf{Line search and convergence.}
As discussed in Section~\ref{sec:convergence}, line search with CCD is crucial for the convergence of Newton iterations while satisfying penetration-free constraints. However, for small time step sizes restricted by the CFL condition, the Newton method can usually converge by itself, given the small displacement and the reference configuration's proximity to the root. To investigate the necessity of line search, we run the same experiment as shown in Figure \ref{fig:validation} (c) (top), involving a droplet colliding with a row of particles. Contact between the droplet and the particles initiates from the 32nd frame. The convergence behavior of Newton iterations was tested with and without using line search at three different CFL numbers (0.7, 1.2, 2). The maximum number of Newton iterations required to converge at each frame is recorded, as shown in Figure \ref{fig:ls_conv}. The maximum number of Newton iterations is capped at 30. From the plot, we observe that line search is unnecessary at CFL=0.7, and convergence of Newton is reached at all steps. At CFL=1.2, the Newton method by itself fails to converge at certain steps (exceeding maximum iterations), but we observe no visible interpenetration artifacts throughout the simulation. At CFL=2.0, the Newton method alone fails immediately after contact, accompanied by obvious interpenetration. Line search is thus a must to ensure the convergence at these two CFL conditions.

\subsection{Three-Dimensional Examples}
\subsubsection{Fluid-shell interaction.}
In this set of examples, we demonstrate interactions between liquids and codimension-1 thin shells across multiple scales, surface tensions, and shell stiffnesses. The scenarios involve large solid deformations, complex fluid topology changes, and delicate interfacial contact, showcasing the robustness and versatility of our method.

\begin{figure}[h]
 \centering
 \includegraphics[width=.48\textwidth]{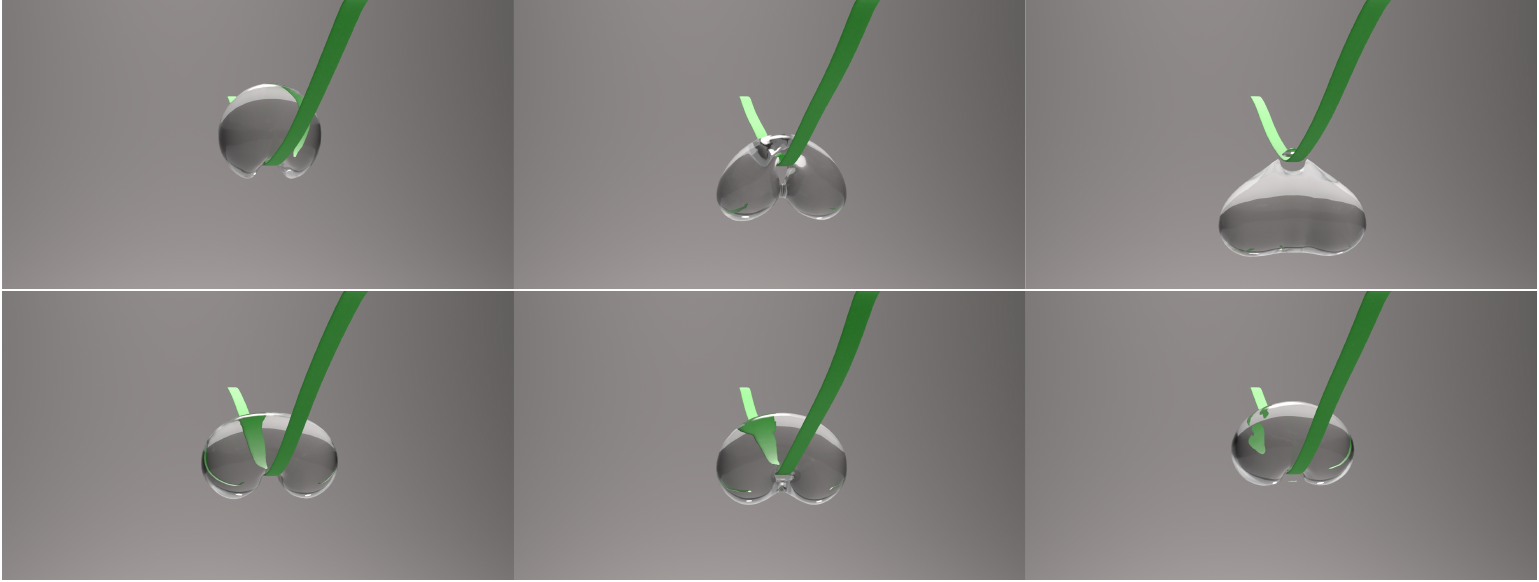}
 \caption{Comparison of liquid topology changes when water drops with low surface tension (top) and high surface tension (bottom) fall on a thin elastic band. Notice the presence of a central void when the band is surrounded by liquid.} 
 \label{fig:drop_on_band}
\end{figure}

\textbf{Basic scenario test.}
This experiment examines a droplet colliding with a thin elastic band (Figure~\ref{fig:drop_on_band}). A low surface tension droplet deforms, reconnects beneath the band forming an elliptical void, and eventually breaks under gravity, leaving a small residue on the band. In contrast, a high surface tension droplet undergoes similar topological changes but remains suspended above. No interpenetration is observed in either case.

\begin{figure}[h]
\centering
 \includegraphics[width=0.48\textwidth]{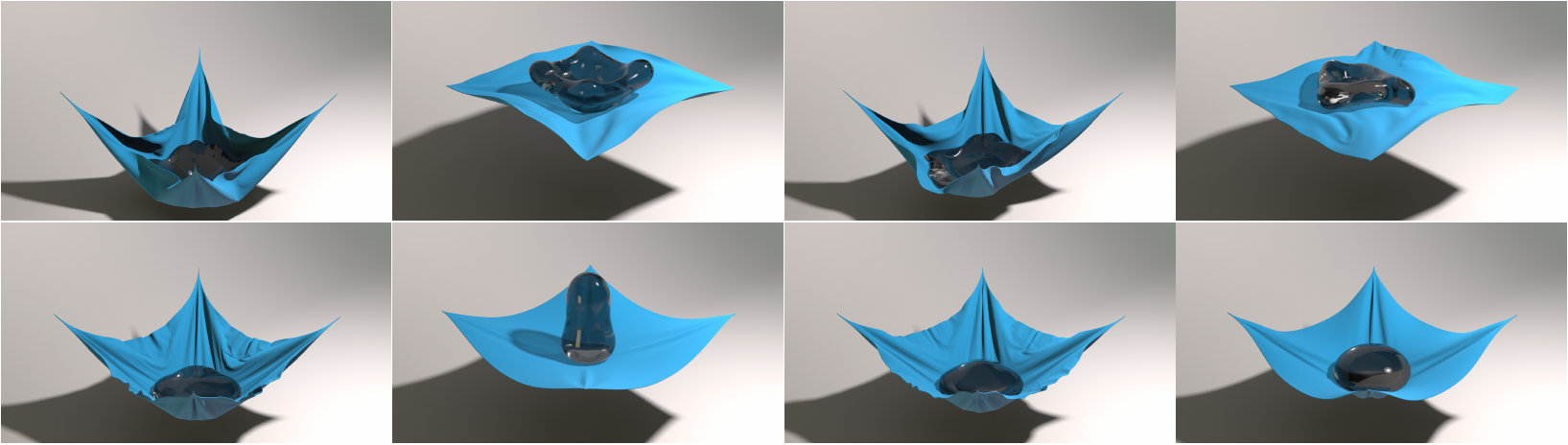}
 \caption{Droplets with varying surface tension bounce on a soft cloth. Top: Low surface tension liquid without damping applied to the cloth. Bottom: High surface tension liquid with Rayleigh damping applied to the cloth.}
 \label{fig:drop_on_cloth}
\end{figure}

\textbf{Bouncing.}
This example demonstrates energy preservation in our method. In the top row of Figure~\ref{fig:drop_on_cloth}, a low surface tension droplet falls onto a soft cloth and repeatedly rebounds, sustaining more than five cycles with minimal energy loss. In the bottom row, higher surface tension and Rayleigh damping lead the system to gradually stabilize, as expected.

\begin{figure}[h]
\centering
 \includegraphics[width=0.48\textwidth]{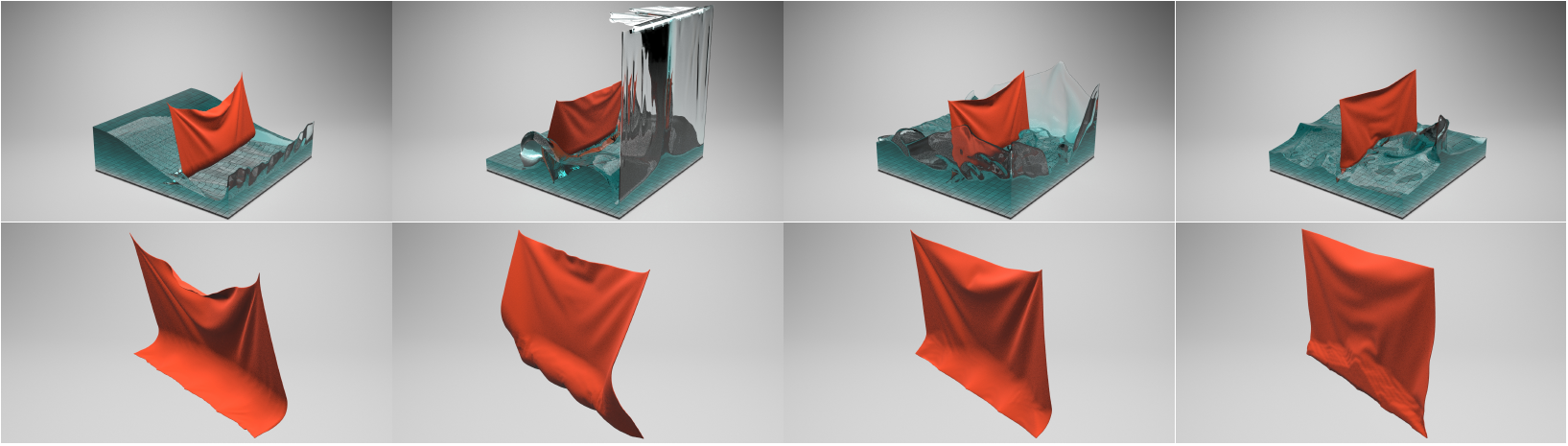}
 \caption{Dam break intercepted by a non-permeable cloth, illustrating the contrast in fluid motion on each side. The cloth is fixed at the two top corners
and five equally spaced vertices at the bottom. The second row showcases intricate surface wrinkles as the cloth interacts with the surrounding
liquids.}
 \label{fig:dam_break}
\end{figure}

\textbf{Dam break.}
While the first two examples involve small droplets, Figure~\ref{fig:dam_break} shows a large tank collapsing over an obstructing shell with narrow gaps. Despite strong impact and severe deformation, the shell remains leak-proof, and the fluid flows smoothly around it. The flow is clearly separated by the shell, and the second row highlights fine wrinkles and tight solid–fluid contact.

\begin{figure}[h]
\centering
 \includegraphics[width=0.48\textwidth]{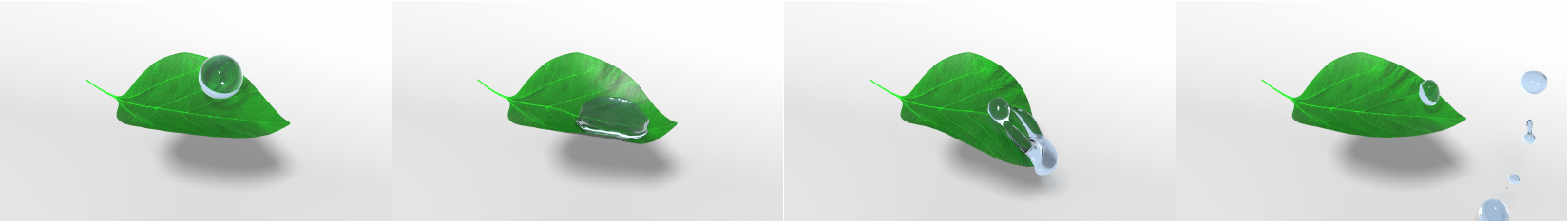}
 \caption{A raindrop splashes down, causing deformation of a hydrophobic leaf with high bending stiffness. The raindrop subsequently fractures into multiple smaller droplets of varying scales, which then rebound into the air.}
 \label{fig:leaf}
\end{figure}

\textbf{Splashing.}
Droplet impingement on solid substrates is governed by the liquid properties, impact velocity, and the mechanical stiffness of the solid. In Figure~\ref{fig:leaf}, high impact velocity, surface tension, and bending stiffness produce rapid, transient contact with significant deformation and topology changes.

\begin{figure}[h]
\centering
 \includegraphics[width=.5\textwidth]{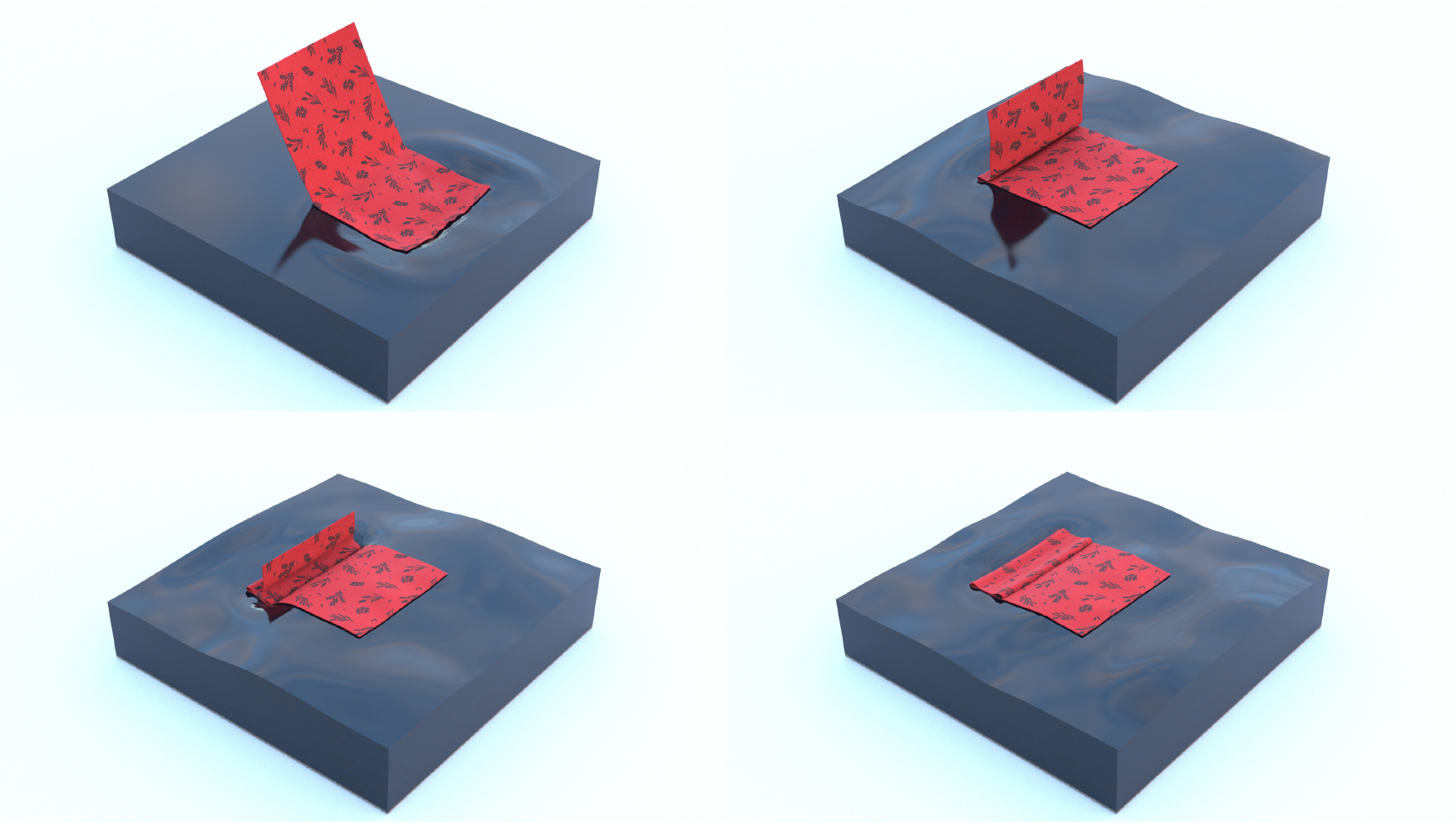}
 \caption{A light, soft cloth initially hung by one side falls onto a tank of water and then floats after being released. Observe the close contact without any permeation and the synchronized motion between the cloth and the underlying surface waves.}
 \label{fig:cloth_on_tank}
\end{figure}

\textbf{Drifting.}
Unlike previous cases with strong impacts, Figure~\ref{fig:cloth_on_tank} shows a light cloth gently placed on a static water tank, with damping approximating air drag. Despite minimal contact forces, the cloth floats stably without penetration. The synchronized deformation of the cloth and fluid surface demonstrates accurate contact handling.

\begin{figure}[H]
\centering
 \includegraphics[width=0.5\textwidth]{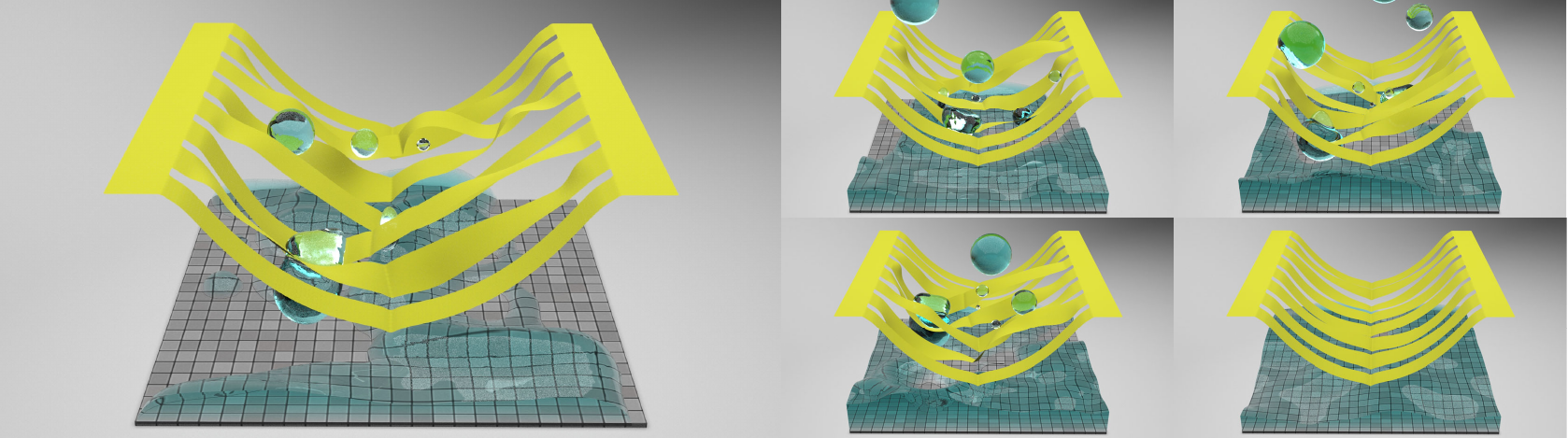}
 \caption{Multiple water drops of varying sizes descend upon a square kirigami structure, featuring a central crease with a rest angle of $110.6 \degree$. To navigate the obstacles, the drops either deform themselves and the kirigami, creating a pathway, or break into smaller fragments.}
 \label{fig:kirigami}
\end{figure}

\textbf{Kirigami.}
To increase geometric complexity, we simulate a V-shaped kirigami with a central crease (Figure~\ref{fig:kirigami}). Droplets of varying scales are randomly poured onto it, exhibiting diverse interactions: some deform to squeeze through gaps, while others fragment to flow around the bands.

\begin{figure}[H]
\centering
 \includegraphics[width=0.5\textwidth]{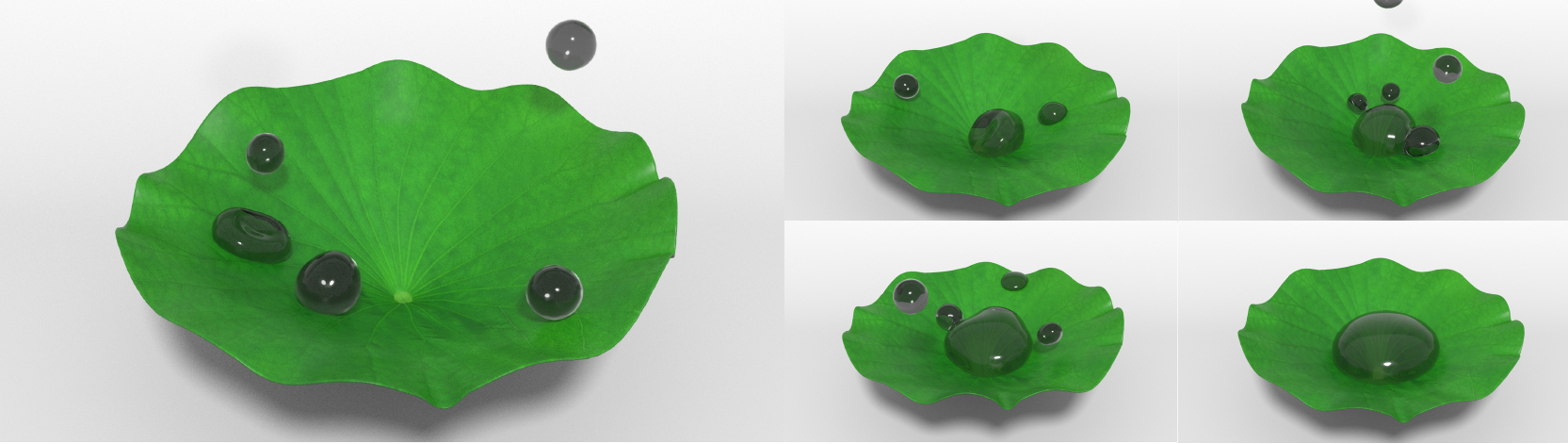}
 \caption{When liquids with surface tension come into contact with a hydrophobic surface like a lotus leaf, they have a tendency to roll off instead of adhering to the surface.}
 \label{fig:lotus}
\end{figure}

\textbf{Lotus leaf.}
The lotus effect describes the self-cleaning behavior of lotus leaves, where water droplets collect dirt and roll off the surface. In Figure~\ref{fig:lotus}, a shell-based lotus leaf interacts with randomly generated raindrops. This example also highlights our method’s strong volume preservation: even with basic level-set advection, each droplet exhibits nearly zero volume change before merging.

\begin{figure}[H]
 \centering
 \includegraphics[width=.48\textwidth]{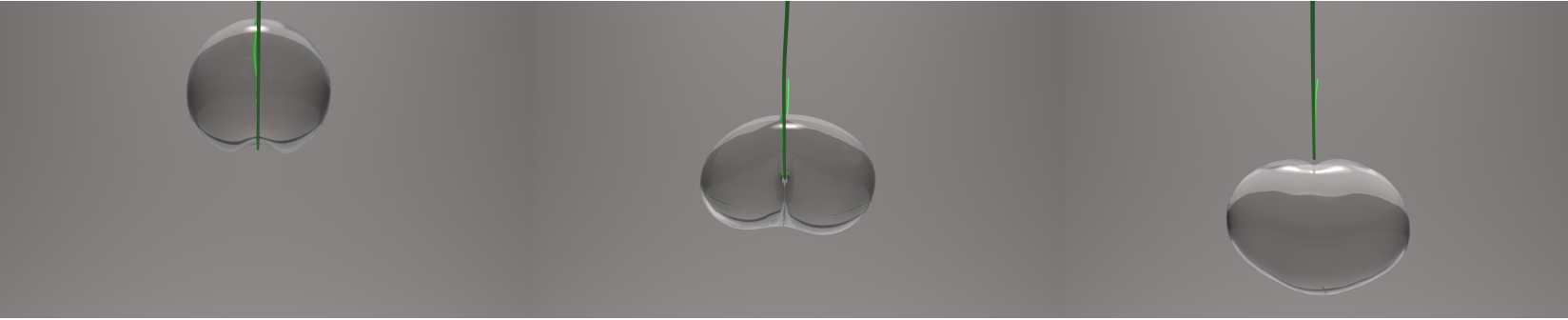}
 \caption{A water drop falls onto an elastic rod and passes through it, showcasing the topological change of the droplet in proximity to the rod.} 
 \label{fig:drop_on_rod}
\end{figure}

\begin{figure}[H]
 \centering
 \includegraphics[width=.5\textwidth]{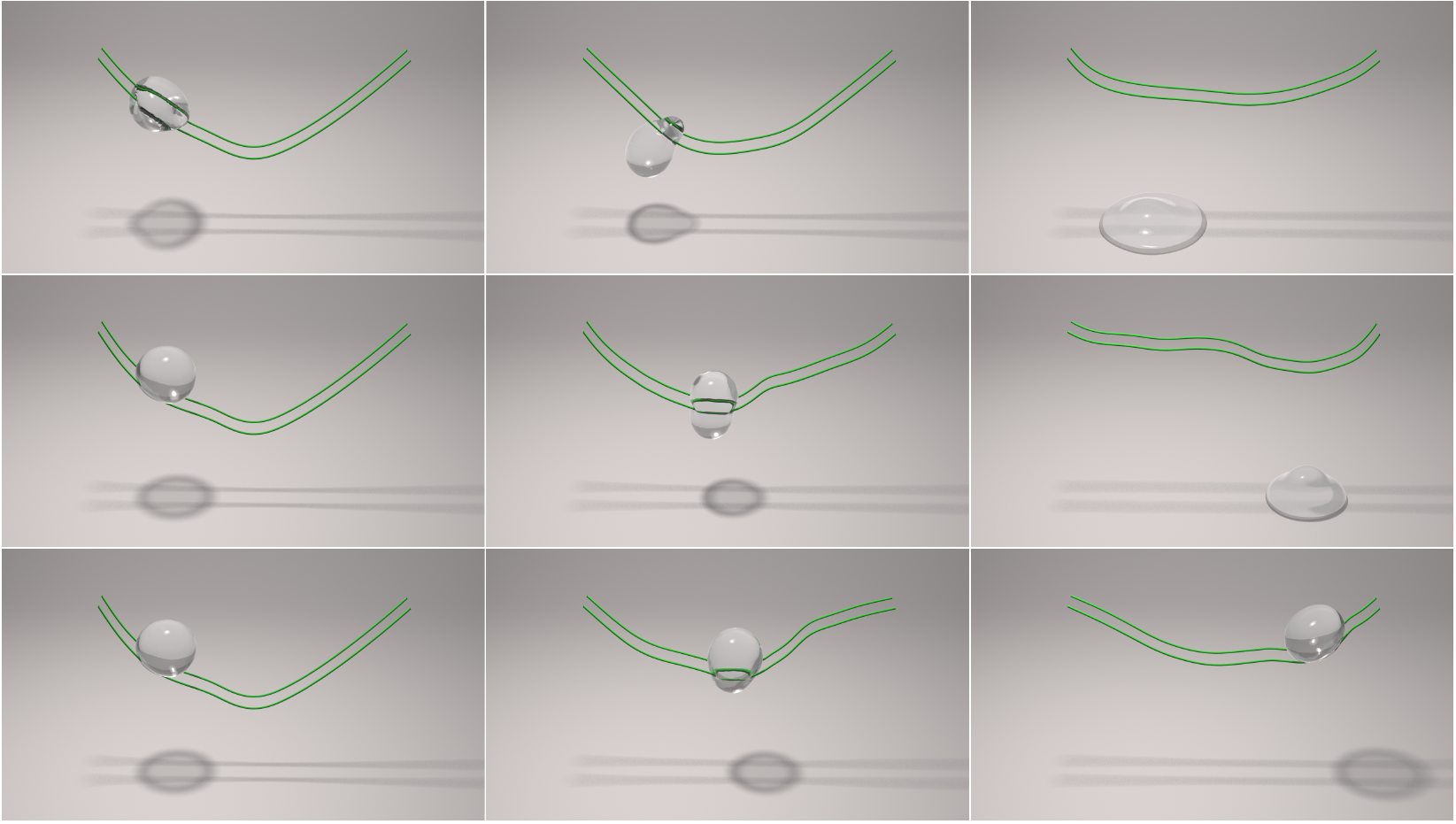}
 \caption{Droplets with varying surface tension slide along a pair of parallel rods. From top to bottom, the surface tension gradually increases.} 
 \label{fig:drop_slide_rod}
\end{figure}

\begin{figure}[H]
 \centering
 \includegraphics[width=.48\textwidth]{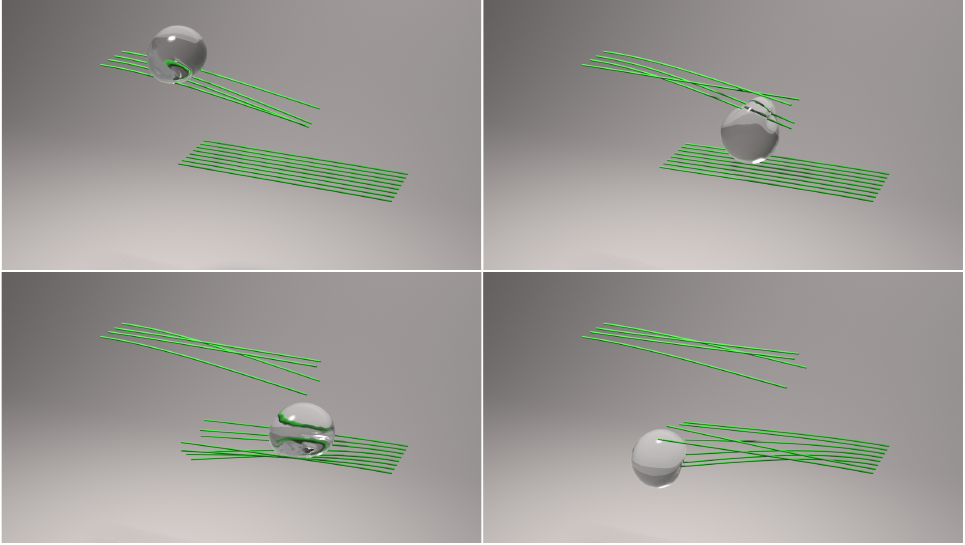}
 \caption{A surface tension droplet glides down two sets of staggered beams made of rigid rods. Note the topological change of the drop shown in the upper right figure as being pierced by the two center rods.} 
 \label{fig:drop_beam}
\end{figure}

\begin{figure}[H]
\centering
 \includegraphics[width=0.48\textwidth]{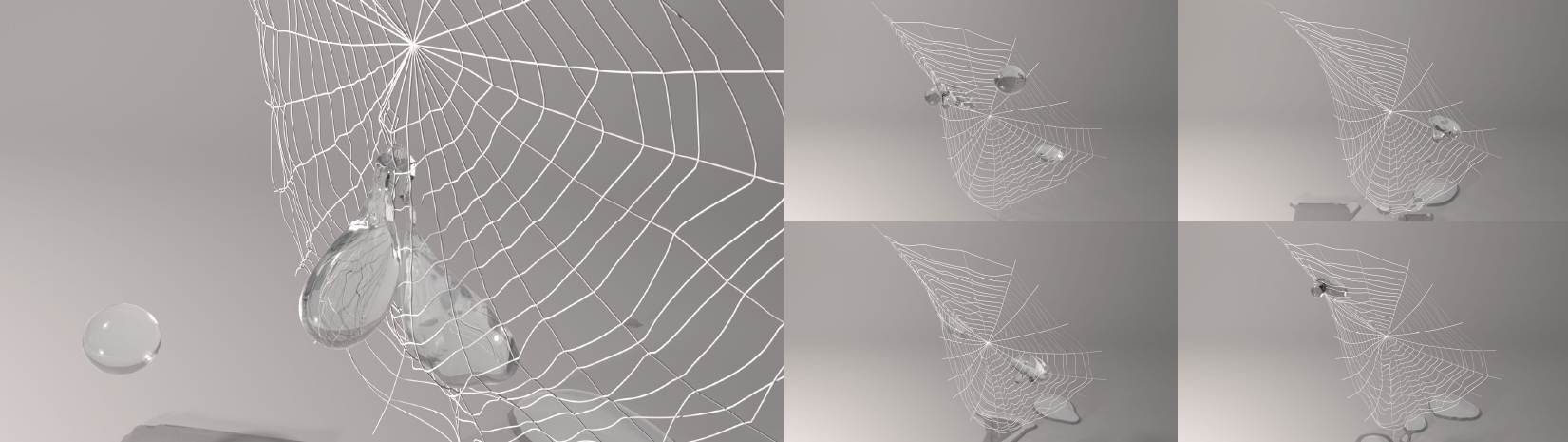}
 \caption{Raindrops descend onto a spider web, slicing through the silks.}
 \label{fig:spiderweb}
\end{figure}
\subsubsection{Fluid-rod interaction}
This set of examples explores interactions between liquids and codimension-2 rods. Although our model excludes viscosity and adhesion, we still demonstrate stable and consistent contact behaviors across varying surface tensions and rod stiffnesses.

\textbf{Basic scenario test.}
Similar to the shell case, we examine a droplet penetrating an elastic rod (Figure~\ref{fig:drop_on_rod}). The process follows three comparable stages, but here the central void forms an extremely thin air tube as the droplet passes through the rod.

\textbf{Sliding.}
To study the effect of surface tension, we simulate a droplet sliding along two parallel rods (Figure~\ref{fig:drop_slide_rod}). As surface tension increases, the droplet travels farther and eventually reaches the other end without falling.

\textbf{Stiff bundles.}
Stiff materials are more prone to penetration, especially at rod tips. In Figure~\ref{fig:drop_beam}, a surface tension droplet slides down two high-stiffness cantilever rods. The drop undergoes topology changes as it is pierced by the central rods, yet remains in close contact with the tips without penetration.

\textbf{Spider web.}
Figure~\ref{fig:spiderweb} shows a soft spider web with non-manifold edges under rainfall, demonstrating our solver’s ability to handle complex rod geometries. Depending on droplet size, impact, surface tension, and material properties, some drops slide along the web while others are split by the silks.

\subsection{Performance}
We provide a summary of the time consumption breakdown in a time step for each of our 3D examples in Table \ref{table:performance}. To optimize performance, most operations, including matrix assembly and grid iteration, are parallelized using OpenMP and oneTBB. Two different solvers are employed for the pressure projection step and the optimization step. For the pressure projection step, we use a standard multigrid-preconditioned conjugate gradient (MGPCG) solver, which solves the Poisson's equation on the grid. Both the building and solving processes are performed on the GPU. The Newton's step is computed using an IDR($s$) solver, preconditioned by a generic incomplete LU factorization with a threshold, provided by the AMGCL library \cite{demidov2019amgcl}. In most examples, we use a small fill-in factor ($p=2$) and a large dropping tolerance ($\tau=10^{-2}$) for fast factorization. However, the leaf (Figure \ref{fig:leaf}) and lotus (Figure \ref{fig:lotus}) examples, which involve large bending stiffness and generate poorly conditioned systems, require different parameter settings. To ensure convergence, we sacrifice factorization speed by retaining more elements. In these cases, we adjust the parameters to $p=18$ and $\tau=10^{-5}$.

The simulations were primarily run on a server with a 32-core CPU and an NVIDIA RTX A4000 graphics card. The lotus and kirigami examples were performed on a personal computer with an 18-core CPU and an NVIDIA RTX A2000 graphics card.

\begin{table}
\centering
\caption{Simulation Setups and Statistics.
(a-b) drops through bands, (c-d) bouncing droplets on cloths, (e) dam break onto a cloth, (f) a drifting cloth on a tank, (g) splashing drops on a leaf, (h) lotus self cleaning, (i) a kirigami in pouring rains, (j) a drop through a rod, (k-m) drops sliding on parallel rods, (n) a drop through a pair of stiff rod bundles, (o) a spider web in pouring rains. $S_{grid}$ and $N_{vert}$ stand for the grid resolution and the number of solid vertices. $t_p$, $t_N$ and $t$ are the time consumptions for pressure projection (sum of two times in each step), optimization, and a complete integral cycle.}
\resizebox{0.4\textwidth}{!}{
\begin{tabular}{lccccccc}
\hline
Case & $S_{grid}$ & $N_{vert}$ & CFL & $t_{P}$(s) & $t_{N}$ (s) & $t$(s)\\
\hline
a. (Fig. \ref{fig:drop_on_band} top) & $160^3$ & 270 & 1 & 0.84 & 1.35 & 3.56 \\
b. (Fig. \ref{fig:drop_on_band} bot) & $160^3$ & 270 & 1 & 0.82 & 1.64 & 4.02\\
c. (Fig \ref{fig:drop_on_cloth} top) & $192^3$ & 12k & 0.7 & 1.03 & 3.7 & 5.95\\
d. (Fig \ref{fig:drop_on_cloth} bot) & $192^3$ & 12k & 0.7 & 1.15 & 4.42 & 7.04\\
e. (Fig \ref{fig:dam_break}) & $192^3$ & 9.2k & 0.7 & 1.55 & 4.9 & 8.03\\
f. (Fig \ref{fig:cloth_on_tank}) & $160^3$ & 6.5k & 0.7 & 0.79 & 2.32 & 4.75\\
g. (Fig \ref{fig:leaf}) & $160^3$ & 17k & 0.5 & 0.71 & 9.2 & 11.6\\
h. (Fig \ref{fig:lotus}) & $192^3$ & 3.6k & 0.5 & 1.26 & 2.68 & 5.44\\
i. (Fig \ref{fig:kirigami}) & $192^3$ & 19.5k & 0.5 & 1.22 & 2.49 & 5.61\\
\hline
j. (Fig \ref{fig:drop_on_rod}) & $160^3$ & 180 & 1 & 0.71 & 0.89 & 2.42\\
k. (Fig \ref{fig:drop_slide_rod} top) & $160^3$ & 360 & 0.7 & 0.75 & 1.14 & 2.68\\
l. (Fig \ref{fig:drop_slide_rod} mid) & $160^3$ & 360 & 0.7 & 0.71 & 0.95 & 2.4\\
m. (Fig \ref{fig:drop_slide_rod} bot) & $160^3$ & 360 & 0.7 & 0.77 & 0.94 & 2.74 \\
n. (Fig \ref{fig:drop_beam}) & $160^3$ & 1.8k & 0.7 & 0.65 & 2.42 & 4.14\\
o. (Fig \ref{fig:spiderweb}) & $192^3$ & 4.2k & 0.7 & 0.98 & 1.13 & 3.1\\ 
\hline
\end{tabular}
}

\label{table:performance}
\end{table}

\subsection{Limitation.}
Our approach has been observed to have several major limitations. Firstly, we only address particle-fluid interpenetration and do not consider collisions between fluids and other types of simplicial complexes. Consequently, penetrations may occur at larger gaps between particles, requiring a relatively dense particle distribution to maintain smooth and leakless contact. Secondly, our incompressibility constraint on fluids considers extra non-fluid cells within a narrowband. This can lead to unphysical volume exchanges when the narrowband cells of two fluid volumes start to overlap, even though the actual interfaces remain isolated. Thirdly, our method only works for hydrophobic interaction and can not handle other contact angles. In addition, our method exhibits numerical dissipation due to the use of semi-Lagrangian advection. Incorporating lower-dissipation advection schemes, such as an advection–reflection solver \cite{jonas2018reflection} combined with unconditionally stable MacCormack correction \cite{selle2008unconditionally}, could help mitigate this issue. Moreover, the Volume-of-Fluid (VoF) method \cite{hirt1981vof} does not naturally integrate into our framework, as accurate curvature estimation remains challenging in VoF formulations, and the computation of solid–fluid distance metrics is also complicated. Finally, the current solver for our optimization problem faces challenges when dealing with stiff solids or special displacement scenarios, such as element inversion. A robust, efficient, and effective factorization approach for the solid submatrix is still lacking, impeding scalability.

\section{Comparisons}
We compare our method with both classic variational solid-fluid coupling approaches \cite{guendelman2005coupling,robinson2008two}, and innovative methods that are recently proposed \cite{ruan2021solid,liu2022hydrophobic,takahashi2022elastomonolith}, to demonstrate the versatility and efficiency of our proposed approach. Coupling particle-based fluids (SPH, MPM, etc) with solids is not considered due to fundamental differences in implementation.

Among the above methods, \cite{liu2022hydrophobic} and \cite{takahashi2022elastomonolith} are limited to volumetric solids and can not be directly extended to thin structures. \cite{ruan2021solid} limits its applications to rigid bodies and necessitates additional data structures (e.g. surface mesh on fluid) and costly geometric operations (e.g. remeshing) for contact handling. \cite{guendelman2005coupling} employs a weak two-way coupling approach, and the customized point-triangle ray casting operation limits its applications to triangle-based shells. This operation is later utilized by \cite{robinson2008two} within a strong two-way coupling framework. Compared to these methods, our proposed framework is naturally applicable to all types of deformables and requires no additional data structures and object-specific operations, as summarized in Table \ref{table:comparison}.

\begin{table}
\centering
\caption{A comparison between our method and previous approaches in terms of supported solid geometries.}
\begin{tabular}{lccc}
\hline
Method & \multicolumn{3}{c}{Solid Support}\\
 (Reference) & Cod-0 & Cod-1 & Cod-2\\
 \hline
\cite{ruan2021solid} & \cmark & \cmark & \xmark \\
\cite{liu2022hydrophobic} & \cmark & \xmark & \xmark \\
\multicolumn{1}{p{2.55cm}}{\cite{takahashi2022elastomonolith}}  & \cmark & \xmark & \xmark \\
\multicolumn{1}{p{2.3cm}}{\cite{guendelman2005coupling}}  & \cmark & \cmark & \xmark \\
\multicolumn{1}{p{2.3cm}}{\cite{robinson2008two}}  & \cmark & \cmark & \xmark \\
Ours & \cmark & \cmark & \cmark \\
\hline
\end{tabular}

\label{table:comparison}
\end{table}

\begin{figure}
 \centering
 \includegraphics[width=.38\textwidth]{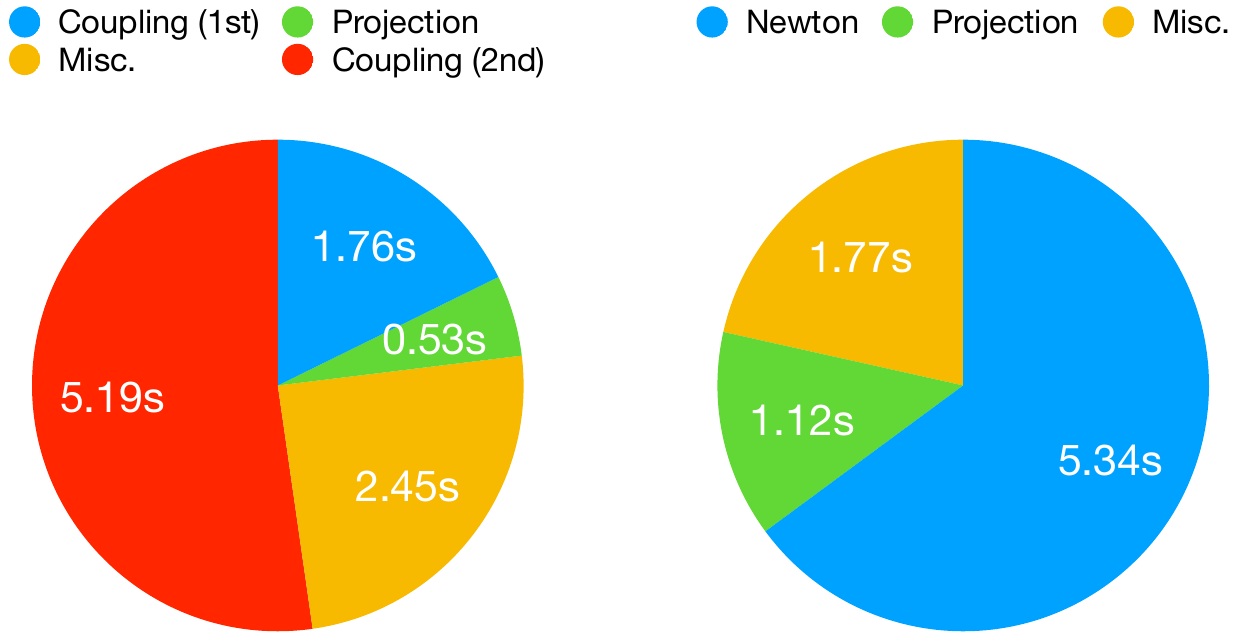}
 \caption{Averaged time cost breakdown per time step for the droplet bouncing example. Left: method proposed by \cite{robinson2008two}; Right: our approach.} 
 \label{fig:pie_chart}
\end{figure}

Performance-wise, we compare our method with \cite{robinson2008two}, the simplest monolithic two-way solid-fluid coupling framework that supports thin shells. Their method involves three implicit solves: two for the indefinite coupling system and one for the pressure projection on the fluid. Our method features a Newton solve for the optimization which requires iteratively solving the coupling system, as well as two pressure solves. The first pressure solve after applying the body force can usually be ignored as in \cite{robinson2008two}. Upon comparing both coupling systems, we observe that the solid parts are identical. The fluid part of \cite{robinson2008two} resembles a Poisson-like structure, while our system's fluid part is mostly diagonal, with limited additional rows and cols corresponds to the number of constraints (isolated fluid domains that needs to preserve their own volumes). The off-diagonal parts, corresponding to the contact, exhibit similar patterns (samples in contact at the interface). Consequently, the factorization speed of the coupling systems should be comparable between the two methods. Additionally, given the relatively small time step sizes constrained by the CFL condition, our Newton solve generally converges within minimal number of iterations. We thus anticipate comparable performance between these two methods. To validate our observation, we ran the droplet bouncing example (Figure \ref{fig:drop_on_cloth}) for 600 frames using both methods with identical setups: surface tension turned off, CFL set to 1, IDR($s$) for coupling solve and MGPCG for Poisson solve. We recorded the total time consumption of linear and Newton solves throughout the simulation, which is then divided by the number of steps to obtain the averaged value, as illustrated in Figure \ref{fig:pie_chart}. Additionally, we recorded the averaged number of iterations per Newton solve, which is 1.93 for this case.

\section{Conclusion And Future Work}
We propose a novel approach for handling contact in the simulation of interactions between Eulerian fluids and codimensional elastic solids. Instead of considering the coupling process in terms of velocity and force, we approach it from a position-level perspective as a continuous trajectory search. Starting from an initial state with interpenetration, our optimizer finds a new configuration that is sufficiently close while ensuring no interpenetration occur. The optimization result is determined collectively by factors such as relative mass, elastic stiffness, fluid incompressibility, and damping.

Through a series of examples, we demonstrate several key benefits of our method. Firstly, it provides robust and accurate contact handling, ensuring immaculate contact between solids and fluids. Secondly, it achieves strict preservation of fluid volume at different scales. Lastly, it minimizes energy dissipation during the coupling process.

We acknowledge that the interactions between solids and fluids are highly nonlinear, especially when multiple terms are considered simultaneously. As such, the use of a single linear system to describe the entire coupling process may have potential limitations in the future. Exploring alternative formulations, such as our proposed approach, could prove valuable in addressing these challenges.

We propose two future directions to further advance our work. Firstly, on the fluid side, there are still some force terms that are currently ignored or treated explicitly, such as viscosity and surface tension. Incorporating these terms into our optimization energy target would make our coupling system more complete. Secondly, on the solid side, a natural progression is to extend our framework to handle multibody systems consisting of codimensional solids, (articulated) rigid bodies, and even multiphase fluids. This would enable a more comprehensive simulation of complex interactions in a wide range of scenarios.

\section*{Acknowledgements}

We express our gratitude to the anonymous reviewers for their insightful feedback. Georgia Tech authors acknowledge NSF IIS \#2433322, ECCS \#2318814, CAREER \#2433307, IIS \#2106733, OISE \#2433313, and CNS \#1919647 for funding support. C. Jiang acknowledges support from NSF 2153851, TRI, Sony, Nvidia, Snap, Style3D, and Disney. M. Li acknowledges support from CMU and Genesis AI. We credit the Houdini education license for video animations.

%% file: author_info.tex
\begin{IEEEbiography}[{\includegraphics[width=0.9in]{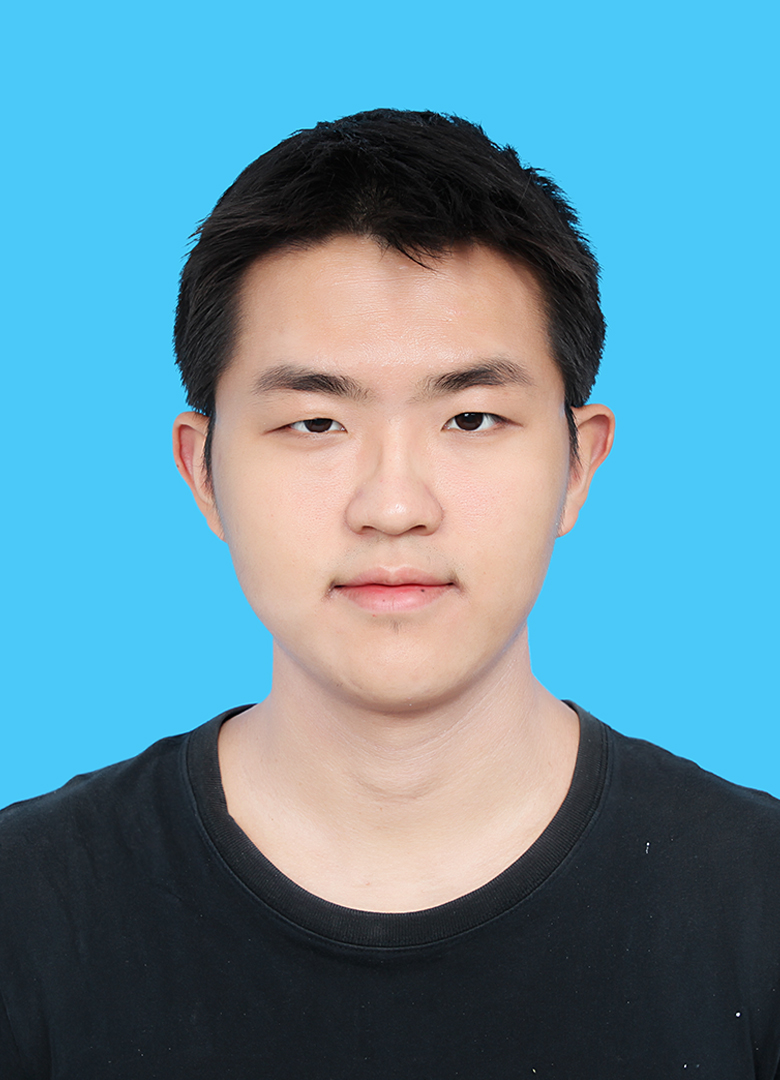}}]{Yuchen Sun}
is Ph.D. student in the School of Interactive Computing at Georgia Tech, advised by Prof. Bo Zhu. He obtained his B.Sc. in Computer Science from Peking University in 2022. His research interest lies in computer graphics, especially in physical simulation. His main focus is to develop high-performance and high-fidelity numerical algorithms to solve computational challenges associated with various kinds of physical systems.
\end{IEEEbiography}

\begin{IEEEbiography}[{\includegraphics[width=0.9in]{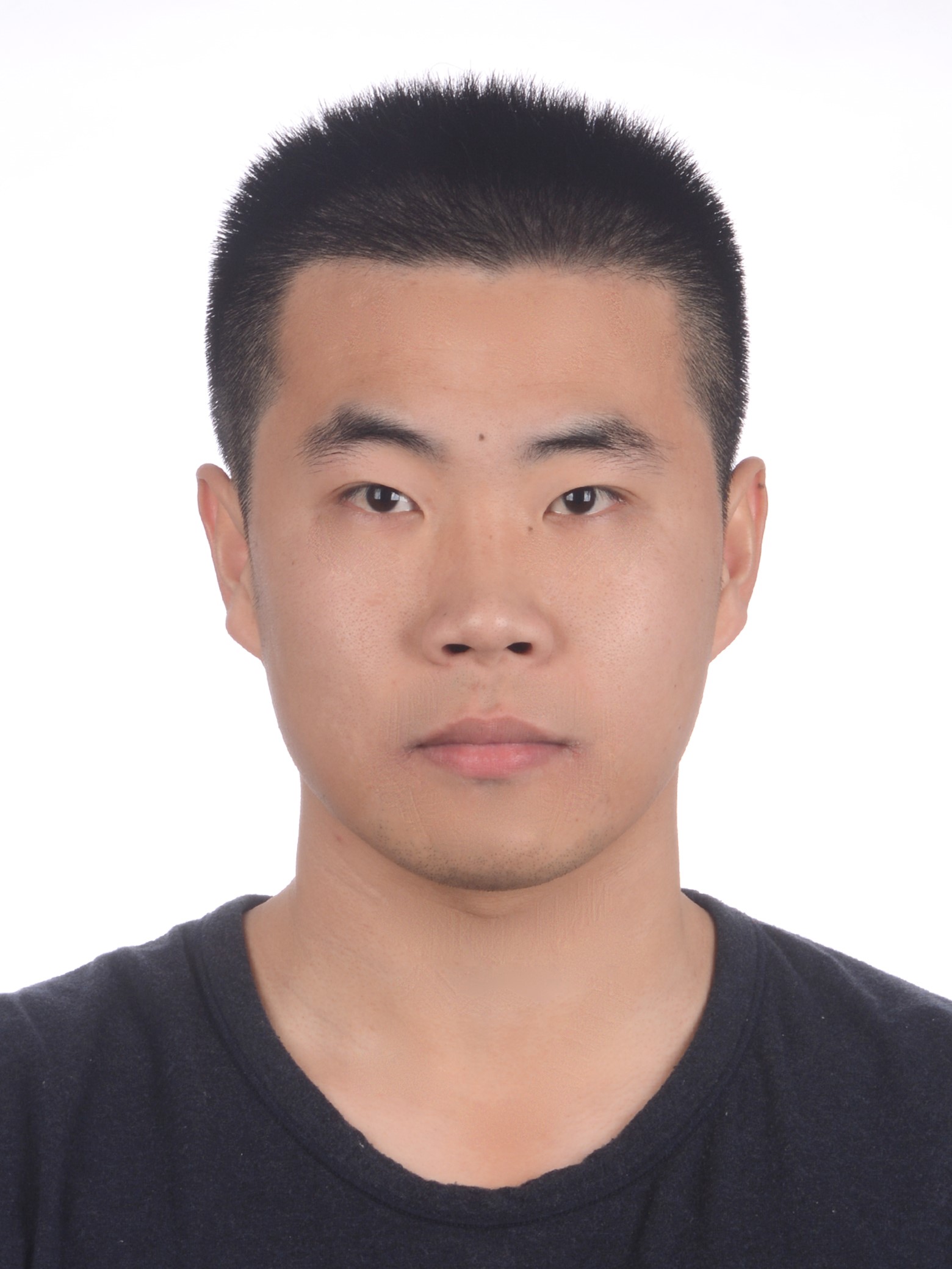}}]{Jinyuan Liu}
received his Ph.D. in Computer Science from Dartmouth College, advised by Prof. Bo Zhu. His disseration is Simulating Cross-Scale Solid-Fluid interaction Phenomena. He obtained his M.S degree in aeronautics \& astronautics in 2017 from Stanford. His research is focused on physically-based modeling and multi-physics simulation, including fluid, deformable bodies, solid-fluid interaction and topology optimization.
\end{IEEEbiography}

\begin{IEEEbiography}[{\includegraphics[width=0.9in]{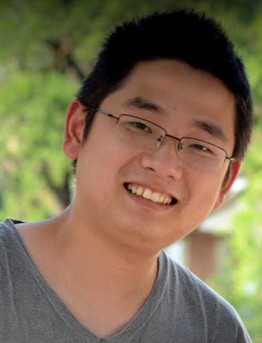}}]{Yin Yang}
is an associate professor with School of Computing, University of Utah. Yin received his Ph.D. from University of Texas at Dallas (with
David Daniel fellowship). He is a recipient of the NSF CRII award (2015) and CAREER award (2019). Yin’s research aims to develop efficient
and customized computing methods for challenging problems in Graphics, Simulation, Machine Learning, Vision, Visualization, Robotics, Medicine, and many other applied areas.
\end{IEEEbiography}

\begin{IEEEbiography}[{\includegraphics[width=0.9in]{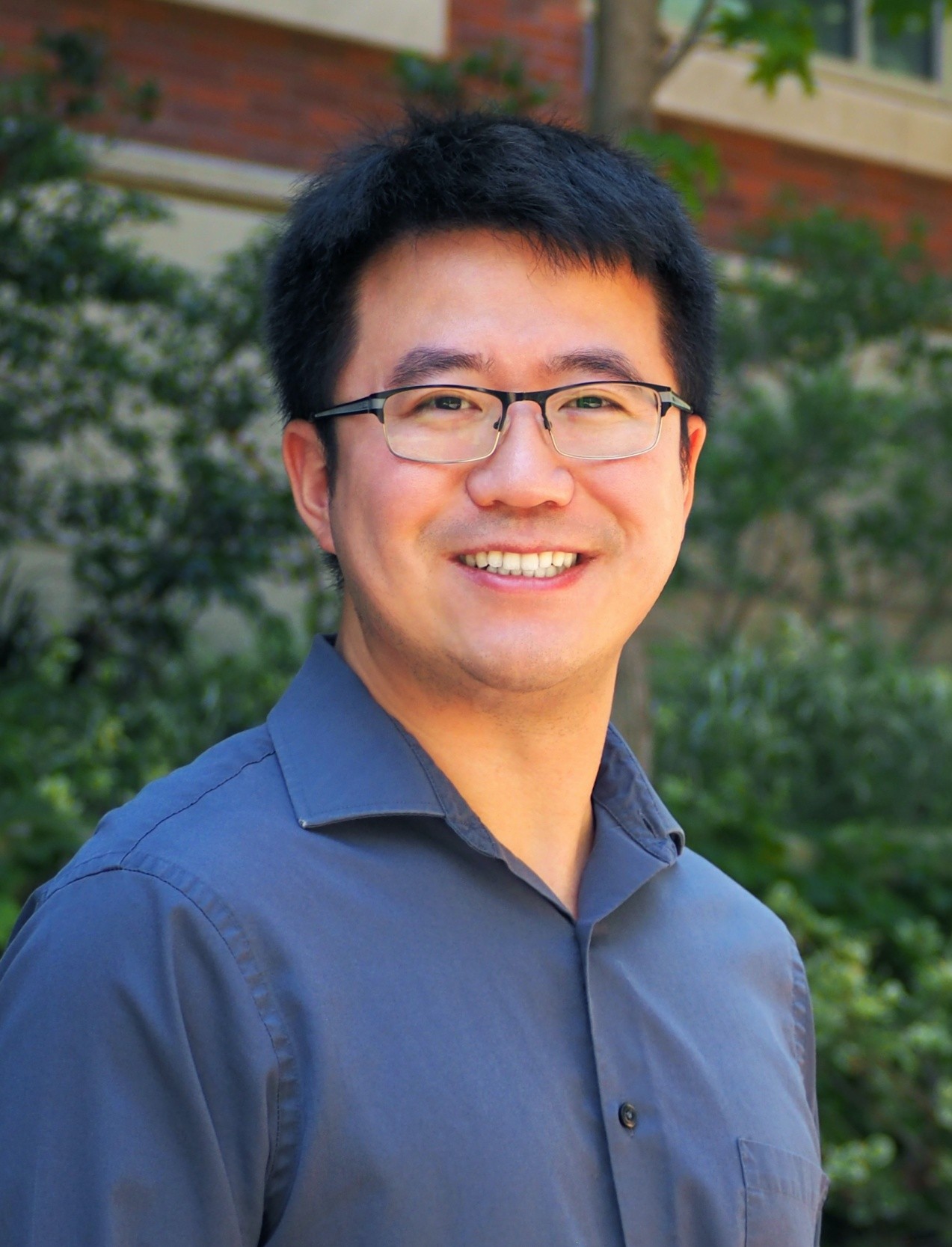}}]{Chenfanfu Jiang}
is a professor of Mathematics at UCLA. He obtained his Ph.D. degree in 2015 from UCLA co-advised by Demetri Terzopoulos and Joseph Teran.
He is a recipient of the UCLA Edward K. Rice Outstanding Doctoral Student Award (2015) and NSF CAREER award (2020). He directs UCLA Multi-Physics Lagrangian-Eulerian Simulations Laboratory with projects spanning scientific computing, computer graphics, metaverse, computational mechanics, machine learning.
\end{IEEEbiography}

\begin{IEEEbiography}[{\includegraphics[width=0.9in]{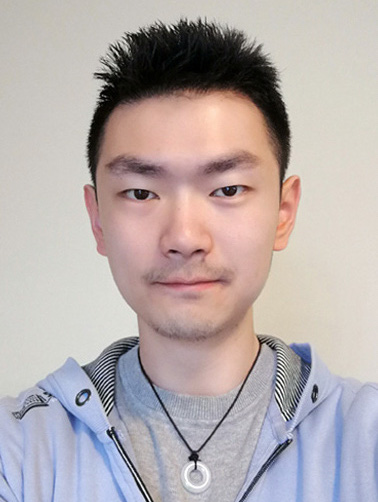}}]{Minchen Li}
is an Assistant Professor in the CS department at Carnegie Mellon University. He was previously an Assistant Adjunct Professor of Mathematics at UCLA following his Ph.D. from the University of Pennsylvania. Minchen’s Ph.D. dissertation, advised by Chenfanfu Jiang, is recognized by the 2021 ACM SIGGRAPH Outstanding Doctoral Dissertation Award for introducing the Incremental Potential Contact (IPC) method, which has led to a series of follow-up works in both academia and industry.
\end{IEEEbiography}

\begin{IEEEbiography}[{\includegraphics[width=0.9in]{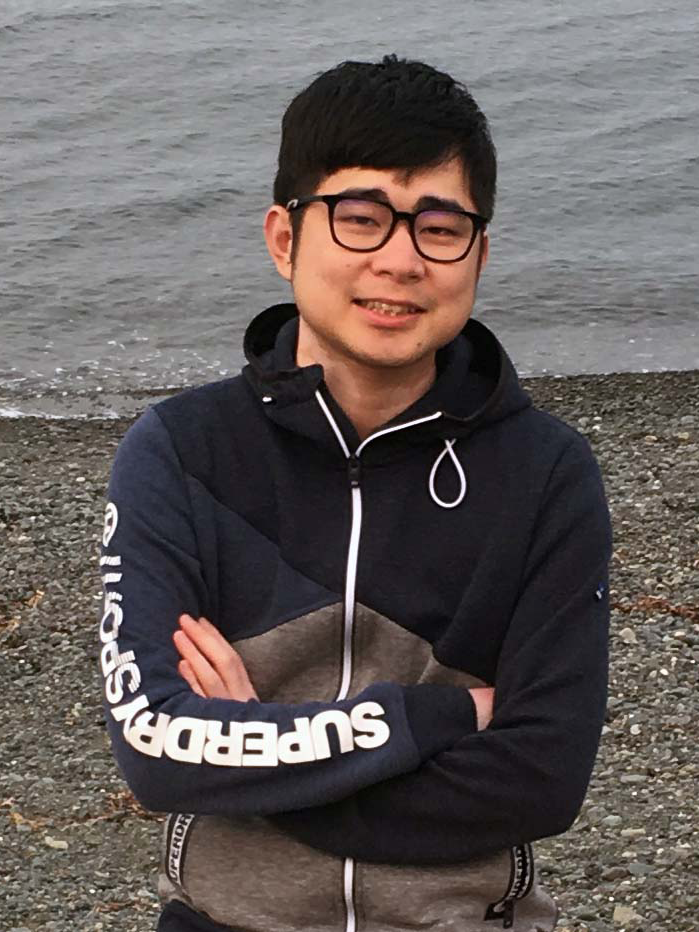}}]{Bo Zhu}
is an Associate Professor in the School of Interactive Computing at Georgia Tech. Prior to that, he was an Assistant Professor of Computer Science at Dartmouth College. He obtained his Ph.D. from Stanford and conducted his postdoctoral research at MIT CSAIL. His research interests include computer graphics, computational physics, computational fluid dynamics, and scientific machine learning. He was awarded the NSF CAREER Award in 2022.
\end{IEEEbiography}

%% file: ref.bib
@article{ruan2021solid,
  title={Solid-fluid interaction with surface-tension-dominant contact},
  author={Ruan, Liangwang and Liu, Jinyuan and Zhu, Bo and Sueda, Shinjiro and Wang, Bin and Chen, Baoquan},
  journal={ACM Transactions on Graphics (TOG)},
  year={2021}
}

@article{robinson2008two,
  title={Two-way coupling of fluids to rigid and deformable solids and shells},
  author={Robinson-Mosher, Avi and Shinar, Tamar and Gretarsson, Jon and Su, Jonathan and Fedkiw, Ronald},
  journal={ACM Transactions on Graphics (TOG)},
  year={2008}
}

@article{batty2007fast,
  title={A fast variational framework for accurate solid-fluid coupling},
  author={Batty, Christopher and Bertails, Florence and Bridson, Robert},
  journal={ACM Transactions on Graphics (TOG)},
  year={2007}
}

@inproceedings{zarifi2017positive,
  title={A positive-definite cut-cell method for strong two-way coupling between fluids and deformable bodies},
  author={Zarifi, Omar and Batty, Christopher},
  booktitle={Proceedings of the ACM SIGGRAPH/Eurographics Symposium on Computer Animation},
  year={2017}
}

@article{da2016surface,
  title={Surface-only liquids},
  author={Da, Fang and Hahn, David and Batty, Christopher and Wojtan, Chris and Grinspun, Eitan},
  journal={ACM Transactions on Graphics (TOG)},
  year={2016}
}

@book{jorge2006numerical,
  title={Numerical optimization},
  author={Jorge, Nocedal and Stephen, J Wright},
  year={2006},
  publisher={Spinger}
}

@article{chen2022simulation,
  title={Simulation and optimization of magnetoelastic thin shells},
  author={Chen, Xuwen and Ni, Xingyu and Zhu, Bo and Wang, Bin and Chen, Baoquan},
  journal={ACM Transactions on Graphics (TOG)},
  year={2022},
  publisher={ACM New York, NY, USA}
}

@article{muller2007position,
  title={Position based dynamics},
  author={M{\"u}ller, Matthias and Heidelberger, Bruno and Hennix, Marcus and Ratcliff, John},
  journal={Journal of Visual Communication and Image Representation},
  year={2007}
}

@article{zheng2009simulation,
  title={Simulation of bubbles},
  author={Zheng, Wen and Yong, Jun-Hai and Paul, Jean-Claude},
  journal={Graphical Models},
  year={2009},
}

@article{zheng2015new,
  title={A new incompressibility discretization for a hybrid particle MAC grid representation with surface tension},
  author={Zheng, Wen and Zhu, Bo and Kim, Byungmoon and Fedkiw, Ronald},
  journal={J. of Computational Physics},
  year={2015},
}

@article{wang2020codimensional,
  title={Codimensional surface tension flow using moving-least-squares particles},
  author={Wang, Hui and Jin, Yongxu and Luo, Anqi and Yang, Xubo and Zhu, Bo},
  journal={ACM Transactions on Graphics (TOG)},
  year={2020},
  publisher={ACM New York, NY, USA}
}

@article{wang2021thin,
  title={Thin-film smoothed particle hydrodynamics fluid},
  author={Wang, Mengdi and Deng, Yitong and Kong, Xiangxin and Prasad, Aditya H and Xiong, Shiying and Zhu, Bo},
  journal={ACM Transactions on Graphics (TOG)},
  year={2021},
  publisher={ACM New York, NY, USA}
}

@book{bridson2015fluid,
  title={Fluid simulation for computer graphics},
  author={Bridson, Robert},
  year={2015},
  publisher={AK Peters/CRC Press}
}

@article{robinson2011symmetric,
  title={A symmetric positive definite formulation for monolithic fluid structure interaction},
  author={Robinson-Mosher, Avi and Schroeder, Craig and Fedkiw, Ronald},
  journal={J. of Computational Physics},
  year={2011},
}

@article{hyde2019unified,
  title={A unified approach to monolithic solid-fluid coupling of sub-grid and more resolved solids},
  author={Hyde, David AB and Fedkiw, Ronald},
  journal={J. of Computational Physics},
  year={2019},
}

@article{gast2015optimization,
  title={Optimization integrator for large time steps},
  author={Gast, Theodore F and Schroeder, Craig and Stomakhin, Alexey and Jiang, Chenfanfu and Teran, Joseph M},
  journal={IEEE transactions on visualization and computer graphics},
  year={2015},
  publisher={IEEE}
}

@article{zsolnai2022flow,
  title={The flow from simulation to reality},
  author={Zsolnai-Feh{\'e}r, K{\'a}roly},
  journal={Nature Physics},
  year={2022},
  publisher={Nature Publishing Group UK London}
}

@article{takahashi2022elastomonolith,
  title={ElastoMonolith: A Monolithic Optimization-Based Liquid Solver for Contact-Aware Elastic-Solid Coupling},
  author={Takahashi, Tetsuya and Batty, Christopher},
  journal={ACM Transactions on Graphics (TOG)},
  year={2022}
}

@inproceedings{stewart2000implicit,
  title={An implicit time-stepping scheme for rigid body dynamics with coulomb friction},
  author={Stewart, David and Trinkle, Jeffrey C},
  booktitle={Proceedings 2000 ICRA. Millennium Conference. IEEE International Conference on Robotics and Automation. Symposia Proceedings (Cat. No. 00CH37065)},
  year={2000},
  organization={IEEE}
}

@article{demidov2019amgcl,
  title={AMGCL: An efficient, flexible, and extensible algebraic multigrid implementation},
  author={Demidov, Denis},
  journal={Lobachevskii Journal of Mathematics},
  year={2019},
  publisher={Springer}
}

@article{zhao2022barrier,
  title={A barrier method for frictional contact on embedded interfaces},
  author={Zhao, Yidong and Choo, Jinhyun and Jiang, Yupeng and Li, Minchen and Jiang, Chenfanfu and Soga, Kenichi},
  journal={Computer Methods in Applied Mechanics and Engineering},
  year={2022},
  publisher={Elsevier}
}

@article{lan2021medial,
  title={Medial IPC: accelerated incremental potential contact with medial elastics},
  author={Lan, Lei and Yang, Yin and Kaufman, Danny and Yao, Junfeng and Li, Minchen and Jiang, Chenfanfu},
  journal={ACM Transactions on Graphics},
  year={2021}
}

@article{lan2022affine,
 author = {Lan, Lei and Kaufman, Danny M. and Li, Minchen and Jiang, Chenfanfu and Yang, Yin},
title = {Affine body dynamics: fast, stable and intersection-free simulation of stiff materials},
year = {2022},
journal = {ACM Transactions on Graphics (TOG)},
}

@article{li2020codimensional,
    author = {Minchen Li and Danny M. Kaufman and Chenfanfu Jiang},
    title = {Codimensional Incremental Potential Contact},
    journal = {ACM Transactions on Graphics (TOG)},
    year = {2021}
}

@article{ferguson2021intersection,
  title={Intersection-free rigid body dynamics},
  author={Ferguson, Zachary and Li, Minchen and Schneider, Teseo and Gil-Ureta, Francisca and Langlois, Timothy and Jiang, Chenfanfu and Zorin, Denis and Kaufman, Danny M and Panozzo, Daniele},
  journal={ACM Transactions on Graphics},
  year={2021}
}

@article{xu2014implicit,
  title={Implicit multibody penalty-baseddistributed contact},
  author={Xu, Hongyi and Zhao, Yili and Barbi{\v{c}}, Jernej},
  journal={IEEE transactions on visualization and computer graphics},
  year={2014},
  publisher={IEEE}
}

@article{tang2012continuous,
  title={Continuous penalty forces},
  author={Tang, Min and Manocha, Dinesh and Otaduy, Miguel A and Tong, Ruofeng},
  journal={ACM Transactions on Graphics (TOG)},
  year={2012},
  publisher={ACM New York, NY, USA}
}

@article{takahashi2021frictionalmonolith,
  title={FrictionalMonolith: a monolithic optimization-based approach for granular flow with contact-aware rigid-body coupling},
  author={Takahashi, Tetsuya and Batty, Christopher},
  journal={ACM Transactions on Graphics (TOG)},
  year={2021},
  publisher={ACM New York, NY, USA}
}

@article{mazhar2015using,
  title={Using Nesterov's method to accelerate multibody dynamics with friction and contact},
  author={Mazhar, Hammad and Heyn, Toby and Negrut, Dan and Tasora, Alessandro},
  journal={ACM Transactions on Graphics (TOG)},
  year={2015},
  publisher={ACM New York, NY, USA}
}

@inproceedings{macklin2020primal,
  title={Primal/dual descent methods for dynamics},
  author={Macklin, Miles and Erleben, Kenny and M{\"u}ller, Matthias and Chentanez, Nuttapong and Jeschke, Stefan and Kim, Tae-Yong},
  booktitle={Computer Graphics Forum},
  year={2020},
  organization={Wiley Online Library}
}

@article{renouf2005conjugate,
  title={Conjugate gradient type algorithms for frictional multi-contact problems: applications to granular materials},
  author={Renouf, Mathieu and Alart, Pierre},
  journal={Computer Methods in Applied Mechanics and Engineering},
  year={2005},
  publisher={Elsevier}
}

@inproceedings{redon2002fast,
  title={Fast continuous collision detection between rigid bodies},
  author={Redon, St{\'e}phane and Kheddar, Abderrahmane and Coquillart, Sabine},
  booktitle={Computer graphics forum},
  year={2002},
  organization={Wiley Online Library}
}

@inproceedings{baraff1994fast,
  title={Fast contact force computation for nonpenetrating rigid bodies},
  author={Baraff, David},
  booktitle={Proceedings of the 21st annual conference on Computer graphics and interactive techniques},
  year={1994}
}

@incollection{kaufman2008staggered,
  title={Staggered projections for frictional contact in multibody systems},
  author={Kaufman, Danny M and Sueda, Shinjiro and James, Doug L and Pai, Dinesh K},
  booktitle={ACM SIGGRAPH Asia 2008 papers},
  year={2008}
}

@article{xie2023contact,
    author={Xie, Tianyi and Li, Minchen and Yang, Yin and Jiang, Chenfanfu},
    title={A Contact Proxy Splitting Method for Lagrangian Solid-Fluid Coupling},
    journal = {ACM Trans. Graph. (SIGGRAPH)},
    year = {2023}
}

@article{baraff1991coping,
  title={Coping with friction for non-penetrating rigid body simulation},
  author={Baraff, David},
  journal={ACM SIGGRAPH computer graphics},
  year={1991},
  publisher={ACM New York, NY, USA}
}

@book{brogliato1999nonsmooth,
  title={Nonsmooth mechanics},
  author={Brogliato, Bernard and Brogliato, B},
  year={1999},
  publisher={Springer}
}

@incollection{andrews2022contact,
  title={Contact and friction simulation for computer graphics},
  author={Andrews, Sheldon and Erleben, Kenny and Ferguson, Zachary},
  booktitle={ACM SIGGRAPH 2022 Courses},
  year={2022}
}

@article{teng2016eulerian,
  title={Eulerian solid-fluid coupling},
  author={Teng, Yun and Levin, David IW and Kim, Theodore},
  journal={ACM Transactions on Graphics (TOG)},
  year={2016},
  publisher={ACM New York, NY, USA}
}

@article{yang2021clebsch,
  title={Clebsch gauge fluid},
  author={Yang, Shuqi and Xiong, Shiying and Zhang, Yaorui and Feng, Fan and Liu, Jinyuan and Zhu, Bo},
  journal={ACM Transactions on Graphics (TOG)},
  year={2021}
}

@article{chen2022unified,
  title={A unified newton barrier method for multibody dynamics},
  author={Chen, Yunuo and Li, Minchen and Lan, Lei and Su, Hao and Yang, Yin and Jiang, Chenfanfu},
  journal={ACM Transactions on Graphics (TOG)},
  year={2022},
  publisher={ACM New York, NY, USA}
}

@incollection{bergou2008discrete,
  title={Discrete elastic rods},
  author={Bergou, Mikl{\'o}s and Wardetzky, Max and Robinson, Stephen and Audoly, Basile and Grinspun, Eitan},
  booktitle={ACM SIGGRAPH 2008 papers},
  year={2008}
}

@inproceedings{grinspun2003discrete,
  title={Discrete shells},
  author={Grinspun, Eitan and Hirani, Anil N and Desbrun, Mathieu and Schr{\"o}der, Peter},
  booktitle={Proceedings of the 2003 ACM SIGGRAPH/Eurographics symposium on Computer animation},
  year={2003},
  organization={Citeseer}
}

@article{li2020incremental,
  title={Incremental potential contact: intersection-and inversion-free, large-deformation dynamics.},
  author={Li, Minchen and Ferguson, Zachary and Schneider, Teseo and Langlois, Timothy R and Zorin, Denis and Panozzo, Daniele and Jiang, Chenfanfu and Kaufman, Danny M},
  journal={ACM Transactions on Graphics (TOG)},
  year={2020}
}

@article{sethian2003level,
  title={Level set methods for fluid interfaces},
  author={Sethian, James A and Smereka, Peter},
  journal={Annual review of fluid mechanics},
  year={2003}
}

@inproceedings{fedkiw2001visual,
  title={Visual simulation of smoke},
  author={Fedkiw, Ronald and Stam, Jos and Jensen, Henrik Wann},
  booktitle={Proceedings of the 28th annual conference on Computer graphics and interactive techniques},
  year={2001}
}

@inproceedings{bridson2002robust,
  title={Robust treatment of collisions, contact and friction for cloth animation},
  author={Bridson, Robert and Fedkiw, Ronald and Anderson, John},
  booktitle={Proceedings of the 29th annual conference on Computer graphics and interactive techniques},
  year={2002}
}

@article{fei2018multi,
  title={A multi-scale model for simulating liquid-fabric interactions},
  author={Fei, Yun and Batty, Christopher and Grinspun, Eitan and Zheng, Changxi},
  journal={ACM Transactions on Graphics (TOG)},
  year={2018},
  publisher={ACM New York, NY, USA}
}

@article{fei2017multi,
  title={A multi-scale model for simulating liquid-hair interactions},
  author={Fei, Yun and Maia, Henrique Teles and Batty, Christopher and Zheng, Changxi and Grinspun, Eitan},
  journal={ACM Transactions on Graphics (TOG)},
  year={2017},
  publisher={ACM New York, NY, USA}
}

@article{akinci2012versatile,
  title={Versatile rigid-fluid coupling for incompressible SPH},
  author={Akinci, Nadir and Ihmsen, Markus and Akinci, Gizem and Solenthaler, Barbara and Teschner, Matthias},
  journal={ACM Transactions on Graphics (TOG)},
  year={2012},
  publisher={ACM New York, NY, USA}
}

@article{fang2020iq,
  title={IQ-MPM: an interface quadrature material point method for non-sticky strongly two-way coupled nonlinear solids and fluids},
  author={Fang, Yu and Qu, Ziyin and Li, Minchen and Zhang, Xinxin and Zhu, Yixin and Aanjaneya, Mridul and Jiang, Chenfanfu},
  journal={ACM Transactions on Graphics (TOG)},
  year={2020},
  publisher={ACM New York, NY, USA}
}

@article{hu2018moving,
  title={A moving least squares material point method with displacement discontinuity and two-way rigid body coupling},
  author={Hu, Yuanming and Fang, Yu and Ge, Ziheng and Qu, Ziyin and Zhu, Yixin and Pradhana, Andre and Jiang, Chenfanfu},
  journal={ACM Transactions on Graphics (TOG)},
  year={2018},
  publisher={ACM New York, NY, USA}
}

@inproceedings{takahashi2019geometrically,
  title={A Geometrically Consistent Viscous Fluid Solver with Two-Way Fluid-Solid Coupling},
  author={Takahashi, Tetsuya and Lin, Ming C},
  booktitle={Computer Graphics Forum},
  year={2019},
  organization={Wiley Online Library}
}

@inproceedings{lu2016two,
  title={Two-way coupling of fluids to reduced deformable bodies},
  author={Lu, Wenlong and Jin, Ning and Fedkiw, Ronald},
  booktitle={Proceedings of the ACM SIGGRAPH/Eurographics Symposium on Computer Animation},
  year={2016}
}

@incollection{narain2010free,
  title={Free-flowing granular materials with two-way solid coupling},
  author={Narain, Rahul and Golas, Abhinav and Lin, Ming C},
  booktitle={ACM SIGGRAPH Asia 2010 papers},
  pages={1--10},
  year={2010}
}

@article{guendelman2005coupling,
  title={Coupling water and smoke to thin deformable and rigid shells},
  author={Guendelman, Eran and Selle, Andrew and Losasso, Frank and Fedkiw, Ronald},
  journal={ACM Transactions on Graphics (TOG)},
  year={2005}
}

@article{zhu2014codimensional,
  title={Codimensional surface tension flow on simplicial complexes},
  author={Zhu, Bo and Quigley, Ed and Cong, Matthew and Solomon, Justin and Fedkiw, Ronald},
  journal={ACM Transactions on Graphics (TOG)},
  year={2014},
  publisher={ACM New York, NY, USA}
}

@article{xing2022position,
  title={Position-Based Surface Tension Flow},
  author={Xing, Jingrui and Ruan, Liangwang and Wang, Bin and Zhu, Bo and Chen, Baoquan},
  journal={ACM Transactions on Graphics (TOG)},
  year={2022},
  publisher={ACM New York, NY, USA}
}

@article{liu2022hydrophobic,
  title={Hydrophobic and Hydrophilic Solid-Fluid Interaction},
  author={Liu, Jinyuan and Wang, Mengdi and Feng, Fan and Tang, Annie and Le, Qiqin and Zhu, Bo},
  journal={ACM Transactions on Graphics (TOG)},
  year={2022},
  publisher={ACM New York, NY, USA}
}

@article{zhu2005animating,
  title={Animating sand as a fluid},
  author={Zhu, Yongning and Bridson, Robert},
  journal={ACM Transactions on Graphics (TOG)},
  year={2005},
  publisher={ACM New York, NY, USA}
}

@article{ihmsen2014sph,
  title={SPH fluids in computer graphics},
  author={Ihmsen, Markus and Orthmann, Jens and Solenthaler, Barbara and Kolb, Andreas and Teschner, Matthias},
  year={2014},
  publisher={The Eurographics Association}
}

@article{jiang2015affine,
  title={The affine particle-in-cell method},
  author={Jiang, Chenfanfu and Schroeder, Craig and Selle, Andrew and Teran, Joseph and Stomakhin, Alexey},
  journal={ACM Transactions on Graphics (TOG)},
  year={2015},
  publisher={ACM New York, NY, USA}
}

@incollection{wojtan2011liquid,
  title={Liquid simulation with mesh-based surface tracking},
  author={Wojtan, Chris and M{\"u}ller-Fischer, Matthias and Brochu, Tyson},
  booktitle={ACM SIGGRAPH 2011 Courses},
  year={2011}
}

@article{macklin2013position,
  title={Position based fluids},
  author={Macklin, Miles and M{\"u}ller, Matthias},
  journal={ACM Transactions on Graphics (TOG)},
  year={2013}
}

@article{setaluri2014spgrid,
  title={SPGrid: A sparse paged grid structure applied to adaptive smoke simulation},
  author={Setaluri, Rajsekhar and Aanjaneya, Mridul and Bauer, Sean and Sifakis, Eftychios},
  journal={ACM Transactions on Graphics (TOG)},
  year={2014},
  publisher={ACM New York, NY, USA}
}

@article{selle2008unconditionally,
  title={An unconditionally stable MacCormack method},
  author={Selle, Andrew and Fedkiw, Ronald and Kim, Byungmoon and Liu, Yingjie and Rossignac, Jarek},
  journal={Journal of Scientific Computing},
  year={2008},
  publisher={Springer}
}

@inproceedings{mcadams2010parallel,
  title={A Parallel Multigrid Poisson Solver for Fluids Simulation on Large Grids.},
  author={McAdams, Aleka and Sifakis, Eftychios and Teran, Joseph},
  booktitle={Symposium on Computer Animation},
  year={2010}
}

@article{enright2002hybrid,
  title={A hybrid particle level set method for improved interface capturing},
  author={Enright, Douglas and Fedkiw, Ronald and Ferziger, Joel and Mitchell, Ian},
  journal={Journal of Computational physics},
  year={2002},
  publisher={Elsevier}
}

@article{hyde2020implicit,
  title={An implicit updated lagrangian formulation for liquids with large surface energy},
  author={Hyde, David AB and Gagniere, Steven W and Marquez-Razon, Alan and Teran, Joseph},
  journal={ACM Transactions on Graphics (TOG)},
  year={2020},
  publisher={ACM New York, NY, USA}
}

@article{monaghan2005smooth,
year = {2005},
author = {J J Monaghan},
title = {Smoothed particle hydrodynamics},
journal = {Reports on Progress in Physics},
}

@phdthesis{liu2024crossscale,
  author  = {Jinyuan Liu},
  title   = {Simulating Cross-Scale Solid-Fluid Interaction Phenomena},
  school  = {Dartmouth College},
  year    = {2024},
  address = {Hanover, NH, USA},
  type    = {Ph.D. dissertation}
}

@article{tampubolon2017multi,
author = {Tampubolon, Andre Pradhana and Gast, Theodore and Kl\'{a}r, Gergely and Fu, Chuyuan and Teran, Joseph and Jiang, Chenfanfu and Museth, Ken},
title = {Multi-species simulation of porous sand and water mixtures},
year = {2017},
address = {New York, NY, USA},
volume = {36},
number = {4},
journal = {ACM Trans. Graph.}
}

@inproceedings{patkar2013hybrid,
author = {Patkar, Saket and Aanjaneya, Mridul and Karpman, Dmitriy and Fedkiw, Ronald},
title = {A hybrid Lagrangian-Eulerian formulation for bubble generation and dynamics},
year = {2013},
booktitle = {Proceedings of the 12th ACM SIGGRAPH/Eurographics Symposium on Computer Animation}
}

@article{lesser2022loki,
author = {Lesser, Steve and Stomakhin, Alexey and Daviet, Gilles and Wretborn, Joel and Edholm, John and Lee, Noh-Hoon and Schweickart, Eston and Zhai, Xiao and Flynn, Sean and Moffat, Andrew},
title = {Loki: a unified multiphysics simulation framework for production},
year = {2022},
journal = {ACM Trans. Graph.},
}

@inproceedings{xuan2024duo,
author = {Li, Xuan and Li, Minchen and Han, Xuchen and Wang, Huamin and Yang, Yin and Jiang, Chenfanfu},
title = {A Dynamic Duo of Finite Elements and Material Points},
year = {2024},
booktitle = {ACM SIGGRAPH 2024 Conference Papers}
}

@article{jonas2018reflection,
author = {Zehnder, Jonas and Narain, Rahul and Thomaszewski, Bernhard},
title = {An advection-reflection solver for detail-preserving fluid simulation},
year = {2018},
journal = {ACM Trans. Graph.}
}

@article{hirt1981vof,
title = {Volume of fluid (VOF) method for the dynamics of free boundaries},
journal = {Journal of Computational Physics},
year = {1981},
author = {C.W Hirt and B.D Nichols}
}
